\documentclass[10pt]{article}
\usepackage{graphicx}
\usepackage{epsfig}

\def\LA{\langle} 
\def\RA{\rangle} 
\def\B<#1|{\mathopen{\LA}{#1}\mathclose{|}} 
\def\K|#1>{\mathopen{|}{#1}\mathclose{\RA}} 
\def\BK<#1|#2>{\mathopen{\LA}{#1}\mathbin{|}{#2}\mathclose{\RA}} 
\def\ME<#1|#2|#3>%
  {\mathopen{\LA}{#1}\mathbin{|}{#2}\mathbin{|}{#3}\mathclose{\RA}} 
\def\KroneckerDelta(#1,#2){\delta_{#1,#2}}
\renewcommand{\vec}[1]{{\mathbf #1}}

\title{$^6$He$+^6$He clustering of $^{12}$Be in a microscopic algebraic approach}

\author{G.~F.~Filippov$^{1)}$, Yu.~A.~Lashko$^{1)}$, S.~V.~Korennov$^{1,2)}$,
K.~Kat\=o$^{3)}$\\{\footnotesize  $^{1)}$Bogolyubov Institute for
Theoretical Physics, 14-b Metrolohichna Street, Kiev-143,
Ukraine}\\{\footnotesize $^{2)}$Physique Nucl\'{e}aire
Th\'{e}orique et Physique Math\'{e}matique,}\\
{\footnotesize Universit\'{e} Libre de Bruxelles, Brussels -1050,
Belgium}\\ {\footnotesize $^{3)}$Graduate School of Science,
Hokkaido University, Sapporo 060-0810, Japan}}

\date{\today}

\begin{document}

\maketitle

\begin{abstract}
The norm kernel of the $A=12$ system composed of two $^6$He
clusters, and the $L=0$ basis functions (in the $SU(3)$ and
angular momentum-coupled schemes) are analytically obtained in the
Fock--Bargmann space. The norm kernel has a diagonal form in the
former basis, but the asymptotic conditions are naturally defined
in the latter one. The system is a good illustration for the
method of projection of the norm kernel to the basis functions in
the presence of $SU(3)$ degeneracy that was proposed by the
authors. The coupled-channel problem is considered in the
Algebraic Version of the resonating-group method, with the
multiple decay thresholds being properly accounted for. The
structure of the ground state of $^{12}$Be obtained in the
approximation of zero-range nuclear force is compared with the
shell-model predictions. In the continuum part of the spectrum,
the $S$-matrix is constructed, the asymptotic normalization
coefficients are deduced and their energy dependence is analyzed.

\end{abstract}

\section{Introduction}

In a number of known papers\cite{FH,Sa,Ho} the resonating-group
method (RGM) has been applied to studies of collisions between
light magic nuclei. The fact that it takes a considerable amount
of energy to excite these nuclei simplifies the calculations but
leaves beyond the scope of the studies multi-channel features of
the continuum spectra of compound systems. Meanwhile, these
features appear naturally in the studies of collisions of light
nuclei with open $p$-shell, when even at comparatively low
energies inelastic exit channels are open.

From a theoretician's viewpoint, a relatively simple example of
collision of light nuclei with open $p$-shell is the scattering of
two $^6$He nuclei. Admittedly, at present it is difficult to stage
such an experiment, but continuum states of $^{12}$Be populated at
the intermediate stage of this scattering are of significant
interest. Along with kinematical and dynamical factors, the Pauli
exclusion principle is an important ingredient in the formation of
these states, and it should be taken into account precisely to
understand its role in multi-channel processes. Finally, a
theoretical analysis of the co-existence of open and closed
channels and its influence to the formation of the continuum
spectrum of $^{12}$Be helps in clarifying the significance of the
closed channels in the structure of wave functions in the
continuum.

Experimental studies of the break-up of $^{12}$Be by Freer {\it et
al.}\cite{Freer1},\cite{Freer2} and an investigation of excited
states of this nucleus in the reaction of two-neutron removal in
an exotic $^{14}$Be beam\cite{Saito} show that there are, in the
energy interval between 12 and 25 MeV, states of $^{12}$Be that
decay primarily through $^6$He$+^6$He and $^8$He$+^4$He channels.
Based on the experimental data is an assumption that there are
states in $^{12}$Be with $^6$He$+^6$He cluster structure. This
assumption is supported by the calculations in the antisymmetrized
molecular dynamics\cite{Kanada,Kanada03}, and in a
quasi-microscopic coupled-channel model\cite{Ito} where this decay
channel is dominant.

Both decay channels were considered by Descouvemont {\it et
al.}\cite{Desc}, where cluster states of $^{12}$Be were calculated
in a generator-coordinate model. Having analyzed partial widths of
the resonance states, the authors pointed out a significant mixing
of the cluster configurations.

In this work, we consider two colliding $^6$He nuclei in a
microscopic framework -- that of the algebraic version of the RGM
(AVRGM). The kinematical information is extracted from the norm
kernel constructed from the single-particle orbitals\cite{BB}
which are the kernels of the integral Bargmann
transform\cite{Bargmann}. Thus the norm kernel (Section 2) is
defined in the Fock--Bargmann space, and there it can be expanded
over the map of the oscillator basis.

Calculation of the norm kernels of several nuclear cluster systems
were earlier made by Hecht {\it et al.} (Ref.\cite{Hecht} and
references therein) and Fujiwara {\it et al.}\cite{FH}. Both
groups utilize the Bargmann space technique\cite{Kramer} and the
$SU(3)$-scalar property of the norm kernels. Apart from the most
tractable, so-called alpha-conjugated systems
($A=4n$)\cite{Suzuki}, the most relevant to our case example of
$^6$Li$+^6$Li was considered in Ref.\cite{Hecht}, where the norm
kernel is tabulated. The projection of the kernel to the basis
states required $SU(3)$ Wigner coefficients\cite{akiyama,hecht82}.
If, however, both basis and the norm kernel are known in their
explicit analytical form, the projection can be done without any
complicated $SU(3)$ recoupling\cite{Fewbody}. The case of
degeneration in the $SU(3)$ basis needs a special consideration,
and a way to resolve the degeneracy is shown in this paper
(Section 3).

In Section 4, we discuss the functions of the angular
momentum-coupled ("physical") basis, which is employed to find the
asymptotic behavior of the coefficients of the expansion of the
wave function in the $SU(3)$ basis and to take account of the
different energies of several decay thresholds, including those
with one of the clusters, or both of them, excited. The
relationships between the two bases are established there.

Action of the antisymmetrizer can be reproduced by means of an
effective potential, properties of which are discussed in Section
5. In Section 6 it is shown how the matrix elements of the
Hamiltonian between the basis functions are calculated. A
completely microscopic approach would require the calculation and
projection of the interaction kernel as well, which is a more
tedious task. Instead, to study the dynamics of the system, we
used the approximation of the zero-range nuclear force, when we
simulated the potential with a few matrix elements, with their
values fitted to reproduce some key experimental data (Sections 7
and 8).

\section{Norm kernel of $^6$He+$^6$He}
\label{sec:kernel}

For a detailed discussion on the norm kernels of binary systems
with open $p$-shell clusters in the Fock--Bargmann space, the
reader is referred to our recent paper\cite{Fewbody}. Following
the procedure described there one gets the
translationally-invariant norm kernel of $^6$He$+^6$He in the form
\begin{eqnarray}
\label{kern:n} I = \sum_n I^n(\bar{\vec u}_1, \bar{\vec u}_2,
\bar{\vec R}; \vec u_1^*, \vec u_2^*, \vec R^*).
\end{eqnarray}

A "ket" Fock--Bargmann state of the system depends on two vectors
$\vec u_1$ and $\vec u_2$ reproducing the dynamics of nucleons in
the open $p$-shells of each of the clusters, and a (Jacobi) vector
$\vec R$ describing the relative motion of the clusters. Since the
internal wave function of a $^6$He cluster is assumed to belong to
the irreducible representation (irrep) $(\lambda,\mu)=(2,0)$, the
cluster-internal vectors $\vec u_k$ are frozen to their second
powers. A "bra" state depends on the vectors with an overbar. All
the vectors are complex-valued, with the complex conjugation
denoted with an asterisk ($^*$).

A term $I^n$ is characterized by the number of oscillator quanta
$n$ along the vector $\vec R$. In order to further expand $I^n$
over $SU(3)$-invariant terms, we first write it as a linear
combination of scalar blocks $\Phi_{(\lambda',\mu')\nu'}$, which
are bilinear in Cartesian components of the vectors $\bar{\vec
u}_k$ (and also $\vec u_k^*$) and homogeneous (of $n$th degree)
over components of $\bar{ \vec R}$, $\vec R^*$. These blocks, in
general, are not $SU(3)$-invariant; they are folded from the
states of various $SU(3)$ representations, the most symmetric
(leading) of which is $(\lambda',\mu')$, and the prime ($^\prime$)
is used hereafter to signify that the function or expression under
question does not entirely belong to the given $SU(3)$
representation. Whenever there are two or more blocks with the
same leading representation, an index $\nu'$ is used to
distinguish them. All blocks are invariant with respect to the
operation of conjugation (interchange of the "bra" and "ket"
vectors). Some examples of such blocks were given in
\cite{Fewbody}, where there were no more than three of them for
each of the systems considered. In the present case, there are 20.

We introduce the following shorthand notation for seven
self-conjugate scalars $a_\kappa$,
\begin{eqnarray*}
&a_1 =(\bar{\vec u}_1 \cdot\vec u^*_1) \,\,\,\,\,\,  a_{13} =
([\bar{\vec
u}_1 \times \bar{\vec R}]\cdot[\vec u^*_1 \times \vec R^*])& \\
&a_2 =(\bar{\vec u}_2 \cdot \vec u_2^*) \,\,\,\,\,\, a_{23} =
([\bar{\vec
u}_2 \times \bar{\vec R}]\cdot[\vec u_2^* \times \vec R^*]) &\\
&a_3 =(\bar{\vec R}\cdot \vec R^*) \,\,\,\,\,\, a_{12} =
([\bar{\vec
u}_1 \times \bar{\vec u}_2]\cdot[\vec u_1^* \times \vec u_2^*]) &\\
& a_{123} =  ([\bar{\vec u}_1 \times \bar{\vec u}_2] \cdot
\bar{\vec R} ) ([\vec u_1^* \times \vec u_2^*] \cdot \vec R^*) &
\end{eqnarray*}
which are the eigenfunctions of the reduced second-order Casimir
operator\cite{Fewbody}
\begin{equation}
\label{s6} {\hat G}_2' = (\bar{\vec u}_1 \cdot \nabla_{\bar {\vec
R} })({\bar{\vec R}} \cdot \nabla_{\bar{\vec u}_1}) + (\bar{\vec
u}_2 \cdot \nabla_{\bar {\vec R}})({\bar {\vec R}} \cdot
\nabla_{\bar{\vec u}_2}) + (\bar{\vec u}_1 \cdot \nabla_{\bar{\vec
u}_2})(\bar{\vec u}_2 \cdot  \nabla_{\bar{\vec u}_1}),
\end{equation}
and their $SU(3)$ symmetry indices are:

$(\lambda_i, \mu_i)=(1,0)$, $(\lambda_{ij}, \mu_{ij})=(0,1)$,
$(\lambda_{123}, \mu_{123})=(0,0)$ ($i\ne j=1,2,3$).

Then the 20 blocks can be written in the form
$$
\Phi_{( \lambda', \mu')\nu'} = \prod_{\kappa}
a^{n_{\kappa}}_\kappa, \mbox{ with }  \lambda' = \sum_{\kappa}
n_\kappa \lambda_\kappa, \,\,\,  \mu' = \sum_{\kappa} n_\kappa
\mu_\kappa.
$$
The values of ${n_{\kappa}}$ are listed in Table~1.
\begin{table}[htb]
\begin{center}
\begin{tabular}{cccccccc}
  \hline
$(\lambda',\mu')\nu'$ \,\,\, \big\backslash \,\,\, $\kappa$ & 1 & 2 & 3 & 12 & 13 & 23 & 123\\
\hline
$(n+4,0)$ & 2 & 2 & $n$ & 0 & 0 & 0 & 0\\
\hline
$(n+2,1)1$ & 1 & 1 & $n$ & 1 & 0 & 0 & 0\\
$(n+2,1)2$ & 1 & 2 & $n-1$ & 0 & 1 & 0 & 0\\
$(n+2,1)3$ & 2 & 1 & $n-1$ & 0 & 0 & 1 & 0\\
\hline
$(n,2)1$ & 0 & 0 & $n$ & 2 & 0 & 0 & 0\\
$(n,2)2$ & 0 & 2 & $n-2$ & 0 & 2 & 0 & 0\\
$(n,2)3$ & 2 & 0 & $n-2$ & 0 & 0 & 2 & 0\\
$(n,2)4$ & 1 & 0 & $n-1$ & 1 & 0 & 1 & 0\\
$(n,2)5$ & 0 & 1 & $n-1$ & 1 & 1 & 0 & 0\\
$(n,2)6$ & 1 & 1 & $n-2$ & 0 & 1 & 1 & 0\\
\hline
$(n-2,3)1$ & 0 & 1 & $n-3$ & 0 & 2 & 1 & 0\\
$(n-2,3)2$ & 1 & 0 & $n-3$ & 0 & 1 & 2 & 0\\
$(n-2,3)3$ & 0 & 0 & $n-2$ & 1 & 1 & 1 & 0\\
\hline
$(n-4,4)$ & 0 & 0 & $n-4$ & 0 & 2 & 2 & 0\\
\hline
$(n+1,0)$ & 1 & 1 & $n-1$ & 0 & 0 & 0 & 1\\
\hline
$(n-1,1)1$ & 0 & 0 & $n-1$ & 1 & 0 & 0 & 1\\
$(n-1,1)2$ & 0 & 1 & $n-2$ & 0 & 1 & 0 & 1\\
$(n-1,1)3$ & 1 & 0 & $n-2$ & 0 & 0 & 1 & 1\\
\hline
$(n-3,2)$ & 0 & 0 & $n-3$ & 0 & 1 & 1 & 1\\
\hline
$(n-2,0)$ & 0 & 0 & $n-2$ & 0 & 0 & 0 & 2\\
 \hline
\end{tabular}
\end{center}
\caption{Powers of $a_\kappa$ in different blocks $\Phi_{(
\lambda', \mu') \nu'}$}
 \label{tab0}
\end{table}

The expansion of the norm kernel then reads
\begin{equation}
\label{s3} I =  {1\over2} \sum_n {1\over2!2!n!}\sum_{( \lambda',
\mu') \nu'} \Lambda'_{( \lambda', \mu') \nu'} \, \Phi_{( \lambda',
\mu')\nu' } (\bar{\vec u}_1, \bar{\vec u}_2, \bar{\vec R}; \vec
u_1^*, \vec u_2^*, \vec R^*) ,
\end{equation}
with the coefficients $\Lambda'_{( \lambda', \mu') \nu'}$ shown in
Table 2.

\begin{table}[htb]
\begin{tabular}{cl}
  \hline
  & \\
$( \lambda', \mu') \nu' $ & \hspace{3cm}$\Lambda'_{( \lambda',
\mu') \nu'}$ \\
& \\
 \hline $(n+4,0)$ & $\{ 1 + (-1)^n\} 2^{-1} 3^{-n} \,
    \{ 30 - 3 \cdot 2^{2 + n} + 2 \cdot 3^n  $ \\ & $  +
      n\left( n-1 \right)
       \left( 28 - 2^n + 2\,n\left( n-5 \right)
         \right)  - 20\,\KroneckerDelta(0,n) -
      24\,\KroneckerDelta(2,n) -
      48\,\KroneckerDelta(4,n) \} $ \\
\hline $(n+2,1)1$ & $  2^{-1} 3^{-n} \,  \{ 4 (1-(-1)^n) n^3 +
[(-2)^n-16] n^2 $ \\ &  $+ [ (-1)^n(2^n-12) + 2 (14-2^n) ] n $ \\
&  $ - 4 [ 5-2^n +(-1)^n(10 - 5 \, 2^n + 3^n) ]
 \}$ \\ & $+
  20  \,\KroneckerDelta(0,n) -
  4   \,\KroneckerDelta(1,n) +
  4/3 \,\KroneckerDelta(2,n) -
  8/9 \, \KroneckerDelta(3,n)$ \\
$(n+2,1)2,3$ & $2^{-1} 3^{-(n+3)}\{\, 27\,
       \left( 1 + (-1)^n  \right) \,
      n \left( n-1 \right)
       \left( 2^n + 20n - 4n^2  -40 \right) $ \\ & $ +
      16\,\,3^{2 + n}\,\KroneckerDelta(2,n) +
      64\,\,3^n\,\KroneckerDelta(4,n)  \}
    $ \\
\hline
$(n,2)1$ & $3^{-n} \{ (n^2-3n+1) + (-1)^n [n^2 + (5-2^n)n +
3^n-2^{n+2}+6 ] $ \\ & $
 - 4 \cdot 3^{n}\,\KroneckerDelta(0,n) +
    2 \cdot 3^{n}\,\KroneckerDelta(1,n) -
    4 \cdot 3^{n-2} \,\KroneckerDelta(2,n) \}$\\
$(n,2)2,3$ & $3^{-n} \left( 1 + {\left( -1 \right) }^n
\right)n(n-1)(n-2)(n-3) -
  16/27\,\KroneckerDelta(4,n)$\\
$(n,2)4$ & $3^{-n} \left( 1 + {\left( -1 \right) }^n \right)
n(n-1)
     \left( 32 - 2^{n-1} + 4\,\left( n-5 \right) \,n
       \right)  $ \\ &  $ -
  8/3 \, \KroneckerDelta(2,n) -
  64/27 \, \KroneckerDelta(4,n)$\\
  $(n,2)5,6$ & $3^{-n} n(n-1)
     \left( 6 + {\left( -2 \right) }^{n-1} - 2\,n +
       2\,{\left( -1 \right) }^n\,\left( 1 + n \right)
       \right)  $ \\ &  $ -
  4/3 \,\KroneckerDelta(2,n)+
  8/9 \, \KroneckerDelta(3,n)$\\

\hline $(n-2,3)1,2$ &$- 3^{-n} 2   \, \left( 1 + {\left( -1
\right) }^n \right) \, n(n-1)(n-2)(n-3)  +
  32/27 \, \KroneckerDelta(4,n)$\\
$(n-2,3)3$ & $-  3^{-n}  \left( n-1 \right) \,n\,
       \left( 6 + {\left( -2 \right) }^{n-1} - 2\,n +
         2\,{\left( -1 \right) }^n\,\left( 1 + n \right)
         \right)   $ \\ &  $ +
  4/3 \,\KroneckerDelta(2,n) -
  8/9 \,\KroneckerDelta(3,n) $\\
\hline $(n-4,4)$ & $ 3^{-n} \left( 1 + {\left( -1 \right) }^n
 \right) \,n(n-1)(n-2)(n-3)
      -
  16/27 \, \KroneckerDelta(4,n)$\\
\hline $(n+1,0)$ & $ - 3^{-n} (1-(-1)^n) \,n\,
     \left( 12 - 2^n + 2\,\left(n-3 \right) \,n
       \right)  +
  4\,\KroneckerDelta(1,n) +
  8/9\,\KroneckerDelta(3,n)$\\
\hline $(n-1,1)1$ & $3^{-n} 2 n\,\left( 2 - {\left( -2 \right)
}^{n-1} - n -
       {\left( -1 \right) }^n\,\left( 2 + n \right)
       \right)  -
  2\,\KroneckerDelta(1,n) +
  8/9 \, \KroneckerDelta(2,n)$\\
$(n-1,1)2,3$ & $ 3^{-n} 2 \left( 1 - {\left( -1 \right) }^n
\right) \, n(n-1)(n-2) - 8/9 \, \KroneckerDelta(3,n)$\\
\hline $(n-3,2)$ & $- 3^{-n} 2 \left( 1 - {\left( -1 \right) }^n
\right) \, n(n-1)(n-2) + 8/9 \, \KroneckerDelta(3,n)$ \\
\hline $(n-2,0)$ & $3^{-n} \left( 1 + {\left( -1 \right) }^n
\right) \, n(n-1) - 4/9 \, \KroneckerDelta(2,n)$ \\
  \hline
\end{tabular}
\caption{Coefficients of the expansion of the norm kernel $I^n$
over the blocks $\Phi_{( \lambda', \mu')\nu' }$.}
 \label{tab2}
\end{table}

There are 14 (cf. Table 2 in \cite{Hecht} and discussion there)
$SU(3)$-invariants which can be constructed as linear combinations
of $\Phi_{( \lambda', \mu')\nu' }$,
\begin{equation}
\label{s4} F_{(\lambda,\mu)\nu} = \sum_{( \lambda', \mu') \nu'}
C^{( \lambda', \mu')\nu' }_{(\lambda,\mu) \nu} \Phi_{( \lambda',
\mu')\nu' }.
\end{equation}
These invariants $F_{(\lambda,\mu)\nu}$ are the projections of the
norm kernels to the subspaces of specific irreducible
representations of $SU(3)$ with $\nu$ is the additional index of
degeneracy.

The coefficients $C^{( \lambda', \mu')\nu' }_{(\lambda,\mu) \nu}$
meet the following conditions.
\begin{itemize}
\item The projections $F_{(\lambda,\mu)\nu}$ are
eigenfunctions of (\ref{s6}), i.e.
\begin{equation}
\label{s5} ( {\hat G}_2' \, - g_2'(\lambda,\mu) )\,
F_{(\lambda,\mu)\nu} = 0.
\end{equation}

\item As we are dealing with two identical clusters, there is an
additional symmetry with respect to interchange of the clusters as
a whole. In algebraic terms, it corresponds to the interchange of
the vectors $\bar{\vec u}_1$ and $\bar{\vec u}_2$, and the
inversion of the vector $\bar{\vec R}$. Evidently, the functions
with even number of quanta, $n=2k$, must be symmetric with respect
to the first operation, while those with $n=2k+1$ --
antisymmetric. This adds an additional requirement: the
projections $F_{(\lambda,\mu)\nu}$ are also eigenfunctions of the
last term of ${\hat G}_2'$, i.e., $(\bar{\vec u}_1 \cdot
\nabla_{\bar{\vec u}_2})(\bar{\vec u}_2 \cdot  \nabla_{\bar{\vec
u}_1})$.

\item
Finally, $F_{(\lambda,\mu)\nu}$ are normalized to the
dimensionality of the irreducible representation
$(\lambda,\mu)$\cite{Harvey}:

\begin{equation}
\label{s7} \hspace{-1cm}\int F_{(\lambda,\mu)\nu}(\vec u_1, \vec
u_2, \vec R; \vec u_1^*, \vec u_2^*, \vec R^*) \, d\mu_b =
\dim[\lambda,\mu] = {(\lambda+1)(\mu+1)(\lambda+\mu+2) \over 2},
\end{equation}
where $d\mu_b$ is the Bargmann measure \cite{Fewbody}.
\end{itemize}

The coefficients $C^{( \lambda', \mu')\nu' }_{(\lambda,\mu) \nu}$
are shown in Appendix A. Inverting\footnote{The word "inverting"
is used here in a broad sense, as the matrix of this coefficients
is not square. It can be reduced to the square form, however,
since the coefficients $\Lambda'_{( \lambda', \mu') \nu'}$ are not
all independent, as seen in Table~2.} the matrix of these
coefficients, one can write the $SU(3)$-projected form of the norm
kernel as follows,
\begin{equation}
\label{s8} I^n =   \sum_{(\lambda,\mu)\nu}
\Lambda_{(\lambda,\mu)\nu} \, F_{(\lambda,\mu)\nu}
\end{equation}
where
\begin{equation}
\label{s9} \Lambda_{(\lambda,\mu)\nu} = {1\over2} \sum_{(
\lambda', \mu')\nu' } {1\over2!2!n!}
(C^{-1})^{(\lambda,\mu)\nu}_{( \lambda', \mu')\nu' } \,
\Lambda'_{( \lambda', \mu')\nu' }
\end{equation}
are the eigenvalues of the norm kernel.

\section{Basis states with $L^\pi=0^+$}

In the following, we restrict ourselves with the terms of the norm
kernel containing the basis states with orbital momentum $L=0$ and
positive parity. They belong to four $SU(3)$ representations,
$(2k+4,0),(2k,2),(2k-4,4)$ and $(2k-2,0)$. All of them have
$n=2k$; the $(2k,2)$ representation is two-fold degenerate, and an
additional index $\nu=1,2,$ will be used to label them, as
$(2k,2)\nu$. We then write the norm kernel
\begin{eqnarray}
I_{L=0}=I_{L=0}(\bar{\vec u}_1, \bar{\vec u}_2, \bar{\vec R}; \vec
u_1^*, \vec u_2^*, \vec R^*)
\end{eqnarray}
expanded over its eigenfunctions $\bar{\Psi}^{(\lambda,\mu)\nu}
\equiv \Psi^{(\lambda,\mu)\nu}_{L=0} (\bar{\vec u}_1, \bar{\vec
u}_2, \bar{\vec R})$ and $\Psi^{*(\lambda,\mu)\nu} \equiv
\Psi^{(\lambda,\mu)\nu}_{L=0}( \vec u_1^*, \vec u_2^*, \vec R^*)$.
\begin{eqnarray}
\label{norm:hilb}
I_{L=0}&=&\sum_{k}\left\{\Lambda_{(2k+4,0)}\bar{\Psi}^{(2k+4,0)}
\Psi^{*(2k+4,0)}+ \Lambda_{(2k,2)_1}\bar{\Psi}^{(2k,2)_1}
\Psi^{*(2k,2)_1}  \right. \nonumber \\
&& +\Lambda_{(2k,2)_2}\bar{\Psi}^{(2k,2)_2} \Psi^{*(2k,2)_2}
 + \Lambda_{(2k-4,4)} \bar{\Psi}^{(2k-4,4)} \Psi^{*(2k-4,4)}
 \nonumber \\
&& \left.
 +\Lambda_{(2k-2,0)} \bar{\Psi}^{(2k-2,0)}
\Psi^{*(2k-2,0)}\right\}.
\end{eqnarray}
(Since the functions with $L\neq0$ are not discussed in this work,
their label $L=0$ will be omitted.)

In order to find analytically the basis functions
$\Psi^{(\lambda,\mu)}$ belonging to non-degenerate $SU(3)$
representations as well as the corresponding eigenvalues, it
suffices to project the kernel (\ref{s8}) to the states with
$L=0$. Alternatively, in the Fock--Bargmann space an
orthonormalized basis can be defined {\it a priori} without the
use of the norm kernel. Then, the eigenvalues are found by folding
the norm kernel with the basis functions. In this way, the
diagonalization of the norm kernel in the $SU(3)$-degenerate basis
is simpler, and solution of the eigensystem is reduced to standard
algebraic procedures.

The fact that the functions $\Psi^{(\lambda,\mu)}$ are
orthonormalized with the Bargmann measure is followed by the
identity
\begin{eqnarray}
\label{a2} \int\Psi^{(\lambda,\mu)}(\vec u_1,\vec  u_2, \vec R)\,
I_{L=0} \, \Psi^{(\bar{\lambda},\bar{\mu})} (\bar{\vec
u}^*_1,\bar{\vec u}^*_2,\bar{\vec R}^*) \, d\mu_b \, d\bar{\mu}_b
=\Lambda_{(\lambda,\mu)}\delta_{\lambda,\bar{\lambda}}\delta_{\mu,\bar{\mu}}
\end{eqnarray}
and an equivalent one,
\begin{eqnarray}
\label{a3} \int I_{L=0} \, \Psi^{(\lambda, \mu)} (\vec u_1,\vec
u_2,\vec R)  d\mu_b =\Lambda_{(\lambda,\mu)}
\Psi^{(\lambda,\mu)}(\bar{\vec u}_1,\bar{\vec  u}_2, \bar{\vec
R}).
\end{eqnarray}
But before we can actually use Eqs.(\ref{a2}--\ref{a3}), we must
find the basis functions $\Psi^{(\lambda,\mu)}$.

\subsection{Non-degenerate case}

The basis functions $\Psi^{(\lambda,\mu)}$ constructed {\it a
priori} in the Fock--Bargmann space must meet the following
requirements. They have to be scalar ($L=0$) eigenfunctions of the
reduced Casimir operator ${\hat G}_2'$ with eigenvalues
$g_2'(\lambda,\mu)$. They also have to be symmetric with respect
to permutations of the vectors ${\vec u}_1, {\vec u}_2$, and be
orthonormalized with the Bargmann measure.

At a given $k$, the least symmetric function $\Psi^{(2k-2,0)}$ has
the simplest form
\begin{eqnarray}
\label{a4}
\Psi^{(2k-2,0)}(\vec u_1,\vec u_2,\vec R)= \sqrt{k\over
6(2k+2)!}\,([{\vec u}_1 \times {\vec u}_2] \cdot {\vec R})^2 {\vec
R}^{2k-2},
\end{eqnarray}
(here and below a shorthand notation $\vec X^{2\nu} \equiv (\vec X
\cdot \vec X)^\nu$ is used).
 The scalar triple product here is characterized by its
$SU(3)$ symmetry indices (0,0) and U(3) indices $[1,1,1]$. It
appears as soon as identical nucleons fill up an oscillator shell.
Note that $\Psi^{(2k-2,0)}$ vanishes when either two of the three
vectors are collinear. Bearing in mind the second power of the
triple product, we conclude that the function (\ref{a4}) has a
zero of the sixth order.

The eigenfunctions belonging to the irreducible representations
$(2k-4,4)$ and $(2k+4,0)$ are
\begin{eqnarray} \label{a5}
\Psi^{(2k-4,4)}(\vec u_1,\vec u_2,\vec R) = \sqrt{{3k(k-1)\over
8(4k^2-1)(2k+1)!}}\left\{ [{\vec u}_1 \times {\vec R}]^2[{\vec
u}_2\times {\vec R}]^2{\vec R}^{2k-4} \right. \nonumber \\
\left. -{2\over3} ([{\vec u}_1 \times {\vec u}_2] \cdot {\vec
R})^2{\vec R}^{2k-2}\right\};
\end{eqnarray}
\begin{eqnarray}
\label{nondeg:f1} \Psi^{(2k+4,0)}({\vec u}_1,{\vec u}_2,{\vec
R})={1\over\sqrt{4(2k+5)(2k)!}} \left\{\vec u_1^2 \vec u_2^2 \vec
R^{2k}-{2\over2k+3}[{\vec u}_1\times {\vec u}_2]^2 \vec R^{2k}
\right. \nonumber
\\
 -{2k\over2k+3}\left([{\vec u}_1\times{\vec R}]^2 \vec u^2_2
\vec R^{2k-2}+[{\vec u}_2\times {\vec R}]^2 \vec u^2_1 \vec
R^{2k-2}\right)  \nonumber
\\
 +{4k(k-1)\over(2k+1)(2k+3)}[{\vec u}_1
\times {\vec R}]^2[{\vec u}_2 \times {\vec R}]^2 \vec R^{2k-4} \nonumber \\
\left. +{4k\over(2k+1)(2k+3)}([{\vec u}_1 \times {\vec u}_2] \cdot
{\vec R})^2 \vec R^{2k-2}\right\}.
\end{eqnarray}
Evidently, the higher the $SU(3)$ symmetry of a function is, the
more complex form it has. The leading terms of the functions
$\Psi^{(2k-4,4)}$ and $\Psi^{(2k+4,0)}$ define their analytical
behavior: the first function has a zero of order 4, the second
does not have zeros.

It has been found\cite{jmp} that a product of even powers of two
vectors can be written as a superposition of hypergeometric
functions, each having a definite $SU(3)$ symmetry. Now we have a
product of even powers of three vectors and again arrive at
expressions having a hypergeometric structure, but there is a
dependence on several independent variables. The $SU(3)$ basis
functions are expressible in terms of hypergeometric functions
$_3F_1(\alpha_1,\alpha_2,\alpha_3;\gamma; z_1,z_2,z_3)$, with the
variables
\begin{eqnarray*}
z_1={[{\vec u}_1 \times {\vec u}_2]^2\over{\vec u}_1^2{\vec
u}_2^2},~~ z_2={[{\vec u}_1 \times {\vec R}]^2\over{\vec
u}_1^2{\vec R}^2},~~ z_3={[{\vec u}_2\times {\vec R}]^2\over{\vec
u}_2^2{\vec R}^2}.
\end{eqnarray*}

This hypergeometric function is defined as follows
\begin{eqnarray*}
\lefteqn{_3F_1(\alpha_1,\alpha_2,\alpha_3;\gamma; z_1,z_2,z_3)}&& \\
& =\sum_{m_1=0}^\infty \sum_{m_2=0}^\infty\sum_{m_3=0}^\infty
{(\alpha_1)_{m_1}(\alpha_2)_{m_2}(\alpha_3)_{m_3}\over{(\gamma)_{m_1+m_2+m_3}m_1!m_2!m_3!}}
z_1^{m_1}z_2^{m_2}z_3^{m_3},&
\end{eqnarray*}
where $(\alpha_i)_{m_i}$ is a Pochhammer symbol\cite{Ryzhyk}.


\subsection{Degenerate case}

As shown in Section \ref{sec:kernel}, there are 6 scalar parts
$\Phi_{(2k,2){\bar \nu}}$ of the norm kernel, which have
$(\lambda', \mu')=(2k,2)$ as their leading $SU(3)$ representation.
Hence there are 6 basis functions with $L^\pi=0^+$ in this
representation. Additional requirements of permutational
symmetries are satisfied by four of them, and there are only two
which are linear-independent.

It is convenient\footnote{Later it will be shown that it is these
functions that are the eigenfunctions of the norm kernel at large
values of the number of quanta.} to choose the following two
functions as the orthonormalized with the Bargmann measure,
Pauli-allowed basis states,
\begin{eqnarray}
\chi^{(2k,2)1}({\vec u}_1,{\vec u}_2,{\vec
R})={k+1\over\sqrt{4(2k+3)!}}\cdot\sqrt{{k(2k-1)\over2k^2+k+1}}\left\{[{\vec
u}_1 \times {\vec R}]^2 \vec u^2_2 \vec R^{2k-2} \right. \nonumber
\\ +[{\vec u}_2 \times {\vec R}]^2 \vec u^2_1 \vec R^{2k-2}
 -{4(k-1)\over2k-1}[{\vec u}_1 \times {\vec R}]^2[{\vec u}_2
\times {\vec R}]^2 \vec R^{2k-4} \nonumber \\
\left. -{2\over(2k-1)(k+1)} ([{\vec u}_1 \times {\vec u}_2] \cdot
{\vec R})^2 \vec R^{2k-2}\right\};
\end{eqnarray}
\begin{eqnarray}
\chi^{(2k,2)2}({\vec u}_1,{\vec u}_2,{\vec
R})={\sqrt{(k+1)(2k^2+k+1)\over{2(2k+3)(2k+3)!}}}  \left\{ [{\vec
u}_1 \times {\vec u}_2]^2 \vec R^{2k} \right. \nonumber \\
- {k(2k-1)\over2k^2+k+1}\left([{\vec u}_1\times {\vec R}]^2 \vec
u^2_2 \vec R^{2k-2}+[{\vec u}_2 \times {\vec R}]^2 \vec u^2_1 \vec
R^{2k-2}\right) \nonumber
\\
 +{4k(k-1)\over2k^2+k+1}[{\vec u}_1 \times {\vec R}]^2[{\vec
u}_2 \times {\vec R}]^2 \vec R^{2k-4} \nonumber
\\
\left. -{k(2k-1)\over2k^2+k+1}([{\vec u}_1 \times {\vec u}_2]
\cdot {\vec R})^2 \vec R^{2k-2}\right\}.
\end{eqnarray}

The leading roles in the behavior of these functions are played by
the expressions,
$$[{\vec u}_1\times {\vec u}_2]^2 {\vec R}^{2k}~~\mbox{and}~~
[{\vec u}_1 \times {\vec R}]^2 {\vec u}_2^2 {\vec R}^{2k-2}+
[{\vec u}_2 \times {\vec R}]^2 {\vec u}_1^2 {\vec R}^{2k-2}.$$
Both expressions are symmetric with respect to the permutation of
the vectors ${\vec u}_1$ and ${\vec u}_2$. Besides, each of them
has a zero of the second order.

In order to resolve the $SU(3)$-degeneracy, we first compute the
following integrals,
\begin{eqnarray}
\label{degen-integrals}
\int\chi^{(2k,2)i}({\vec u}_1,{\vec
u}_2,{\vec R}) I_{L=0}\chi^{(2k,2)j} (\bar{\vec u}^*_1,\bar{\vec
u}^*_2,\bar{\vec R}^*) d\mu_b d\bar{\mu}_b=
\lambda_{ij}(k),~i,j=1,2.
\end{eqnarray}
The coefficients $\lambda_{ij}(k)$ are shown in the Appendix B.

At a given $n$ the norm kernel $I^n$ can be written (cf.
Eq.~(\ref{s8})) as a sum of $SU(3)$-projected norm kernels
$I^{(\lambda,\mu)}$. We shall deal with a relevant part of the
norm kernel, $I^{(2k,2)}_{L=0}$ and write it as
\begin{eqnarray}
\label{sep}
I_{L=0}^{(2k,2)}=\sum_{i,j=1}^2\lambda_{ij}(k)\chi^{(2k,2)i}(\bar{\vec
u}_1,\bar{\vec u}_2,\bar{\vec R}) \chi^{(2k,2)j}({\vec
u}^*_1,{\vec u}^*_2,{\vec R}^*)
\end{eqnarray}
It follows from Eqs.~(\ref{a3}) and (\ref{sep}) that
$I_{L=0}^{(2k,2)}$ is a degenerate kernel of the integral equation
(\ref{a3}), hence it can be presented in the form of
Hilbert--Schmidt expansion,
\begin{eqnarray}
\label{deg:kern}
I_{L=0}^{(2k,2)}=\Lambda_{(2k,2)1}(k)\bar{\Psi}^{(2k,2)1}
{\Psi}^{*(2k,2)1}
 + \Lambda_{(2k,2)2}(k) \bar{\Psi}^{(2k,2)2} \Psi^{*(2k,2)2}.
\end{eqnarray}
We shall search the solution of the integral equation in the form
\begin{eqnarray}
\Psi^{(2k,2)1}&=&\cos\alpha(k)\chi^{(2k,2)1}-\sin\alpha(k)\chi^{(2k,2)2};
\nonumber \\
\Psi^{(2k,2)2}&=&\sin\alpha(k)\chi^{(2k,2)1}+\cos\alpha(k)\chi^{(2k,2)2},
\end{eqnarray}
satisfying the norm condition for the functions $\Psi^{(2k,2)1}$
and $\Psi^{(2k,2)2}$. We arrive to a set of linear equations of
the second order for the angle $\alpha(k)$, with the solution
\begin{eqnarray*}
\cos\alpha(k)={1\over\sqrt{2}}\sqrt{1+{1\over\sqrt{1+x^2(k)}}}\,\,,~~
\sin\alpha(k)={1\over\sqrt{2}}\sqrt{1-{1\over\sqrt{1+x^2(k)}}}\,\,,
\end{eqnarray*}
\begin{eqnarray}
x(k)={2\lambda_{12}(k)\over\lambda_{11}(k)-\lambda_{22}(k)} .
\end{eqnarray}
The structure of the basis functions in the $SU(3)$-degenerate
case depends on the number of quanta $2k$ through the angle
$\alpha$ (Table 3, the rightmost column). At $k=3$ (the minimal
number of quanta allowed for this representation) $\alpha$ is
close to $\pi/6$. At $k\rightarrow\infty$, $\alpha$ limits to
zero, because
\begin{eqnarray}
x(k)\rightarrow -{6\sqrt{2}\over
k}\left(1+{4\over3k}\right)\rightarrow 0.
\end{eqnarray}
The determinant of the set is equated to zero, and the eigenvalues
are found from a quadratic equation as
\begin{eqnarray}
\Lambda_{(2k,2)_{1,2}}={\lambda_{11}(k)+\lambda_{22}(k)\over2}\pm
{\lambda_{11}(k)-\lambda_{22}(k)\over2}\sqrt{1+x^2(k)},
\end{eqnarray}
The dependence of the eigenvalues on $k$ is also shown in Table 3.
One of them, $\Lambda_{(2k,2){2}}$, is zero at $k=3$. Therefore at
the minimally allowed number of quanta there is only one function
in the representation $(2k,2)$.

The eigenvalues of the kernel (\ref{sep}) of the integral equation
have a finite limiting point, where it is, therefore, impossible
to uniquely define the eigenfunctions $\Psi^{(2k,2)\nu}$. In this
point, any pair of functions obtained after a unitary
transformation of $\chi^{(2k,2){1,2}}$ would be a solution of the
integral equation. If $k\rightarrow\infty$, then
$\lambda_{12}(k)\rightarrow 0$, and both $\lambda_{11}(k)$ and
$\lambda_{22}(k)$ limit to 1. Meanwhile, at any finite value of
$k$, however small the values of $\lambda_{12}(k)$,
$\lambda_{11}(k)-1$ and $\lambda_{22}(k)-1$ are, the
eigenfunctions $\Psi^{(2k,2)\nu}$ are unique. That is why at large
values of $k$ it is better to work with the {\it limiting
solution} of the integral equation rather than with a solution of
a {\it limiting integral equation}.

\subsection{Eigenvalues of the norm kernel}

Now that the basis functions of irreducible representations of the
$SU(3)$ group are constructed, the eigenvalues of the
non-degenerate states can be computed using Eq.(\ref{a2}). The
non-vanishing eigenvalues are
\begin{eqnarray*}
\Lambda_{(2k+4,0)}=1-{2^{2k}(2k^2-k+6)-4k(2k-1)(2k^2-5k+7)-15\over3^{2k}}
\,\,\,\,\, (k \ge 5)   ;
\\
\Lambda_{(2k-4,4)}=1-{2^{2k}\cdot7-55\over3^{2k}} \,\,\,\,\, (k
\ge 3); ~~\Lambda_{(2k-2,0)}=1-{2^{2k-2}-13\over3^ {2k}}
\,\,\,\,\, (k \ge 2).
\end{eqnarray*}
Note that each of these eigenvalues corresponds to any state with
these $SU(3)$ symmetry indices, not only with $L=0$.

Summarizing our calculations, we present the Hilbert--Schmidt
expansion of the norm kernel $I_{L=0}$, eliminating the vanishing
eigenvalues:
\begin{eqnarray}
\label{norm_kmin} I_{L=0}&=&
\sum_{k=5}\Lambda_{(2k+4,0)}\bar{\Psi}^{(2k+4,0)}\Psi^{*(2k+4,0)}+
\sum_{k=4}\Lambda_{(2k,2)_2}\bar{\Psi}^{(2k,2)_2}\Psi^{*(2k,2)_2}
\nonumber
\\
&&
+\sum_{k=3}\Lambda_{(2k,2)_1}\bar{\Psi}^{(2k,2)_1}\Psi^{*(2k,2)_1}+
\sum_{k=3}\Lambda_{(2k-4,4)}\bar{\Psi}^{(2k-4,4)}\Psi^{*(2k-4,4)}
\nonumber
\\
&&+\sum_{k=2}\Lambda_{(2k-2,0)}\bar{\Psi}^{(2k-2,0)}\Psi^{*(2k-2,0)}.
\end{eqnarray}
We observe here that the minimal number of quanta allowed by the
Pauli principle is $n=2k=4$. The number of allowed states
increases with $k$, and it is only at $n=2k \ge 10$ quanta of
relative motion of the clusters where all possible $SU(3)$
representations are realized in the norm kernel $I_{L=0}$ (cf.
Table 3).

\begin{table}[htb]
\label{t2}
\begin{center}
\begin{tabular}{ccccccc}
\hline $k$ & $\Lambda_{(2k+4,0)}$ & $\Lambda_{(2k,2)_2}$&
$\Lambda_{(2k,2)_1}$ &
$\Lambda_{(2k-4,4)}$ & $\Lambda_{(2k-2,0)}$& $\cos\alpha(k)$ \\
\hline
2 & 0 & 0 & 0  & 0 & 1.1111 &  \\
3 & 0 & 0 & 0.8313  & 0.4609 & 0.9959 & 0.8700 \\
4 & 0 & 0.3117 & 0.9229 & 0.7352 & 0.9922 & 0.8792 \\
5 & 0.2134 & 0.5864 & 0.9592 & 0.8795 & 0.9959 & 0.8896\\
6 & 0.4694 & 0.7657 & 0.9792 & 0.9461 & 0.9981 & 0.9000 \\
7 & 0.6730 & 0.8713 & 0.9896 & 0.9760 & 0.9991 & 0.9102\\
8 & 0.8092 & 0.9307 & 0.9947 & 0.9893 & 0.9996 & 0.9198\\
9 & 0.8926 & 0.9633 & 0.9973 & 0.9953 & 0.9998 & 0.9286\\
10 & 0.9411 & 0.9807 & 0.9986 & 0.9979& 0.9999 & 0.9366\\
 \hline
\end{tabular}
\end{center}
 \caption[]{Eigenvalues $\Lambda_{(\lambda,\mu)}$ of the norm
 kernel for the system $^6$He+$^6$He}
\end{table}

\section{Wave function of $^6$He$+^6$He}

We now have the complete basis\footnote{The basis functions depend
on $k$ through the $SU(3)$ index $\lambda$, but it sometimes will
be shown as a lower index for the sake of clarity; no confusion
should occur.} $\{\Psi^{(\lambda,\mu)\nu}_k\}$ of Pauli-allowed
states with $L=0$ in the channel $^6$He$+^6$He. The wave function
$\Psi^E_{L=0}$ of this channel can be expanded over this basis,
\begin{eqnarray}
\label{d1} \Psi^E_{L=0}=\sum_k \sum_{(\lambda_k,\mu)\nu}
C^{(\lambda,\mu)\nu}_k(E) \,  \Psi^{(\lambda,\mu)\nu}_k,
\end{eqnarray}
where $E$ is the energy of the state counted from the threshold of
the decay of $^{12}$Be into two $^6$He nuclei, and
$C^{(\lambda,\mu)\nu}_k(E)$ are the expansion coefficients to be
found using a set of equations of the AVRGM\cite{AVRGM}.

When the bound states are studied, the utilization of the $SU(3)$
basis poses no problems. The expansion coefficients decrease
rapidly enough to employ a reasonably limited number of basis
states in (\ref{d1}) even if the states are close to the decay
threshold. The dependence of $C^{(\lambda,\mu)\nu}_k(E)$ on $k$
provides some information on the validity of the shell model,
because in the latter only the states with the minimal value of
$k$ are employed, and the cluster degrees of freedom are frozen.

When, however, the continuous states are studied in AVRGM, the use
of the asymptotic values of $C^{(\lambda,\mu)\nu}_k(E)$ at large
$k$ becomes indispensable. Meanwhile, the asymptotic behavior is
known not for these coefficients, but for the coefficients
$C^{(l_1,l_2,l)}_k(E)$ appearing in the expansion of the wave
function over the states of the "physical", angular
momentum-coupled basis $\{\Phi_k^{(l_1,l_2,l)}\}$. The states of
this basis (referred to as "$l$-basis" in what follows) are
labelled by the number of quanta $2k$, angular momenta of each of
the $^6$He clusters $l_1$ and $l_2$, and the angular momentum of
their relative motion $l$. In AVRGM, the asymptotic behavior of
$C^{(l_1,l_2,l)}_k(E)$ is expressed in terms of the Hankel
functions of the first and second kind
$H^{(1,2)}_{l+1/2}(\sqrt{2E}\sqrt{4k+2l+3})$ and the scattering
$S$-matrix elements\cite{Scat}.

The angular momentum-coupled basis for the system $^6$He+$^6$He,
at a given $k$ and $L=0$, consists of five orthonormalized
functions, with the following structure,
$$
\Phi^{(l_1,l_2,l)}_{k}({\vec u}_1,{\vec u}_2,{\vec
R})=N^{(l_1,l_2,l)}_{k}
$$
\begin{eqnarray}
 \times \left\{\{{\vec
u}_1\otimes{\vec u}_1\}_{l_1}\otimes\{{\vec u}_2\otimes{\vec
u}_2\}_{l_2}\otimes\{{\vec R}\otimes...\{{\vec R}\otimes\{{\vec
R}\otimes{\vec R}\}_{l'}\}_{l''}...\}_{l}\right\}_{L=0}.
\end{eqnarray}
Here $\{{\vec A}\otimes{\vec B}\}_{l}$ is an irreducible tensor
product\cite{Varsh} of the rank $l$, $N^{(l_1,l_2,l)}_{k}$ is a
norm factor.

The $l$-basis functions have the form
\begin{eqnarray}
\Phi^{(0,0,0)}_k&=&{1\over6\sqrt{(2k+1)!}} \, {\vec u}_1^2{\vec
u}_2^2{\vec
R}^{2k} \nonumber \\
\Phi^{(2,2,0)}_k&=&{1\over\sqrt{20(2k+1)!}}\left\{ [{\vec u}_1
\times  {\vec u}_2]^2{\vec R}^{2k}-{2\over3} \, {\vec u}_1^2{\vec
u}_2^2{\vec
R}^{2k}\right\} \nonumber \\
\Phi^{(2,2,2)}_k&=&\sqrt{{9k\over 28(2k+3)(2k+1)!}}\left\{ ([{\vec
u}_1 \times {\vec u}_2] \cdot {\vec R})^2{\vec R}^{2k-2}-{1\over3}
[{\vec
u}_1 \times {\vec u}_2]^2{\vec R}^{2k} \right. \nonumber \\
&& \left. -{1\over3}\left([{\vec u}_1 \times {\vec R}]^2{\vec
u}_2^2+ [{\vec u}_2 \times {\vec R}]^2{\vec u}_1^2\right) {\vec
R}^{2k-2}+ {4\over9} \, {\vec u}_1^2{\vec u}_2^2{\vec
R}^{2k}\right\}; \nonumber \\
\Phi^{(2,2,4)}_k&=& \sqrt{{5k(k-1)\over14(2k+3)(2k+5)(2k+1)!}}
\left\{-([{\vec u}_1 \times {\vec u}_2] \cdot {\vec R})^2 {\vec
R}^{2k-2} \right. \nonumber \\
&&+{7\over2} [{\vec u}_1\times {\vec R}]^2[{\vec u}_2 \times {\vec
R}]^2{\vec R}^{2k-4}+ {4\over5}[{\vec u}_1 \times {\vec
u}_2]^2{\vec R}^{2k-2}\nonumber \\
&& \left. -2\left([{\vec u}_1 \times {\vec R}]^2{\vec u}_2^2+
[{\vec u}_2 \times {\vec R}]^2{\vec u}_1^2\right){\vec R}^{2k-2}+
{4\over5}\, {\vec u}_1^2{\vec u}_2^2{\vec R}^{2k}\right\}
\nonumber
\end{eqnarray}
\vspace{-7mm}
\begin{eqnarray}
 {1\over\sqrt{2}} \left(\Phi^{(2,0,2)}_k+\Phi^{(0,2,2)}_k \right)=\sqrt{ k \over 8(2k+3)(2k+1)!}
 \phantom{aaaaaaaaaaaaaaaaaaaaaa}  \nonumber \\
\times \left\{ -\left([{\vec u}_1 \times {\vec R}]^2{\vec u}_2^2+
[{\vec u}_2 \times {\vec R}]^2{\vec u}_1^2\right){\vec R}^{2k-2}+
{4\over3}{\vec u}_1^2{\vec u}_2^2{\vec R}^{2k}\right\}.
\end{eqnarray}

The functions of the $SU(3)$ basis and those of the $l$-basis are
related by a unitary transformation,
\begin{eqnarray}
\label{unit} \Psi^i_k={\cal U}_{ij}(k)\Phi^{j}_k,~~\mbox{where
}\,\,\, \Psi^i_k\equiv\Psi^{(\lambda,\mu)_\nu}_k,~~
\Phi^{j}_k\equiv\Phi^{(l_1,l_2,l)}_{k},~~ i,j= 1, \dots,5.
\end{eqnarray}
As the matrix elements of this transformation has a cumbersome
form at small $k$, we only show (Table 4) their values at
$k\rightarrow\infty$\footnote{These limiting values are used to
find the asymptotic behavior of the expansion coefficients of the
wave function in either basis.}.

\begin{table}[htb]
\begin{tabular}{cccccc}
 \hline
& $\Phi^{(0,0,0)}_k$ & $\Phi^{(2,2,0)}_k$ &
${1\over\sqrt{2}}\left\{\Phi^{(2,0,2)}_k+
\Phi^{(0,2,2)}_k\right\}$ & $\Phi^{(2,2,2)}_k$ & $\Phi^{(2,2,4)}_k$ \\
\hline $\Psi^{(2k-2,0)}$ & ${2\over3\sqrt{3}}$ &
${5\over3\sqrt{15}}$ & $-{2\over3\sqrt{3}}$ &
${\sqrt{14}\over3\sqrt{3}}$& 0\\
$\Psi^{(2k-4,4)}$ & ${2\sqrt{2}\over3\sqrt{3}}$ & $-
{4\sqrt{2}\over3\sqrt{15}}$& $-{2\sqrt{2}\over3\sqrt{3}}$&
$-{4\over3\sqrt{21}}$ &
${\sqrt{3}\over\sqrt{35}}$\\
$\Psi^{(2k,2)1}$ &  ${2\over3}$& ${2\over3\sqrt{5}}$&
${1\over3}$& $-{2\over3}\sqrt{{2\over7}}$ &$-2\sqrt{{2\over35}}$\\
$\Psi^{(2k,2)2}$ &0  &$\sqrt{{2\over5}}$ &0 &$-{1\over\sqrt{7}}$ &
${4\over\sqrt{35}}$\\
$\Psi^{(2k+4,0)}$ & ${1\over3}$ &$-{2\over3\sqrt{5}}$ &
${2\over3}$&
${2\sqrt{2}\over3\sqrt{7}}$ &${2\sqrt{2}\over\sqrt{35}}$\\
 \hline
\end{tabular}
\caption{Matrix of the unitary transformation between the $SU(3)$
basis and the $l$-basis at $k\rightarrow\infty$}
 \label{tab1}
\end{table}
We observe in Table 4 that the states of the $SU(3)$
representations $(2k-4,4)$ è $(2k,2)1$ are dominated by the
$s$-wave of the relative motion of the clusters ($53\%$ in either
state). The $d$-wave is dominant in $\Psi^{(2k+4,0)}$ and
$\Psi^{(2k-2,0)}$ ($67\%$ in the latter state). Finally, $l=4$ is
dominant in the $(2k,2)2$ representation ($46\%$). Note that in
the function $\Psi^{(2k-2,0)}$ the $l=4$ component is absent.

Since all eigenvalues of the norm kernel become equal to $1.0$
when $k\to\infty$, the diagonal form of the expansion
(\ref{norm:hilb}) holds after a unitary transformation of the
basis $\Psi^{(\lambda,\mu)\nu}$. Hence the same unitary
transformation can be used to express the asymptotic behavior of
$C^{(\lambda,\mu)\nu}_k(E)$ in terms of that of
$C^{(l_1,l_2,l)}_k(E)$, which is needed to close the set of AVRGM
equations for the coefficients $C^{(\lambda,\mu)\nu}_k(E)$.

\section{Effective potential}

In contrast with the $SU(3)$ basis\footnote{The antisymmetrizer
$\hat{A}$ and $SU(3)$ generators commute.}, the functions of the
$l$-basis are not eigenfunctions of the antisymmetrizer $\hat{A}$.
Action of $\hat{A}$ to an $l$-basis function does not change the
value of $k$, but still produces a superposition of several basis
functions. Needless to say, those superpositions of $l$-basis
functions which correspond to the functions of the $SU(3)$ basis
are eigenfunctions of $\hat{A}$. Thus the operator $\hat{A}$ can
be represented as a sum of operators $\hat{I}_k$, each acting in
the subspace spanned on the basis functions with the number of
quanta $2k$. The functions of the $SU(3)$ basis are eigenfunctions
of $\hat{I}_k$, whereas the $l$-basis functions are not.

Consider two $SU(3)$ basis functions, which we simply denote here
as $\Psi_1(k)$ and $\Psi_2(k)$, and two functions of the
$l$-basis, $\Phi_1(k)$ and $\Phi_2(k)$. Let $\lambda_i(k)$ be the
eigenvalue of $\hat{I}_k$ corresponding the $\Psi_1(k)$,
\begin{equation}
\hat{I}_k\Psi_i(k)=\lambda_i(k)\Psi_i(k).
\end{equation}
The two sets of basis functions are related through a unitary
transform,
\begin{eqnarray}
\Phi_1(k)&=&\cos\phi(k)\Psi_1(k)+\sin\phi(k)\Psi_2(k) \nonumber \\
\Phi_2(k)&=&-\sin\phi(k)\Psi_1(k)+\cos\phi(k)\Psi_2(k).
\end{eqnarray}
In the $SU(3)$ basis, the operator $\hat{I}_k$ has a diagonal
form, so that
\begin{eqnarray*}
\label{eff} \sum_{k',j'} \left\{ <k,j|{\hat H}|k',j'>
-E\lambda_j(k) \delta_{kk'} \delta_{jj'} \right\} C^{j'}_{k'}=
0,~~{\hat H}={\hat T}+{\hat U},
\end{eqnarray*}
where $C^i_k$ are coefficients of expansion of the wave function
$\Psi$ of the nucleus in this basis,
\begin{eqnarray}
\Psi=\sum_k\left(C^1_k\Psi_1(k)+C^2_k\Psi_2(k)\right).
\end{eqnarray}
Meanwhile, in the $l$-basis the matrix of $\hat{I}_k$ takes the
form
\begin{equation}
\left(\begin{array}{cc}
{\lambda_1(k)+\lambda_2(k)\over2}+{\lambda_1(k)-\lambda_2(k)\over2}\cos2\phi(k)
& -
{\lambda_1(k)-\lambda_2(k)\over2}\sin2\phi(k)\vspace{0.3cm} \\
-{\lambda_1(k)-\lambda_2(k)\over2}\sin2\phi(k) &
{\lambda_1(k)+\lambda_2(k)\over2}-
{\lambda_1(k)-\lambda_2(k)\over2}\cos2\phi(k)
\end{array}\right).
\end{equation}
Eq.(\ref{eff}) can be also written as
\begin{eqnarray}
\label{eff2} \sum_{k',j'}\left\{<k,j|{\hat H}|k',j'>
-E(\lambda_j(k)-1)\delta_{kk'} \delta_{jj'} \right\}
C^{j'}_{k'}=E\,C^{j}_k
\end{eqnarray}
which means that the matrix of the operator of effective potential
appearing due to $\hat{I}_k$ is equal to the difference of the
matrix of the operator $\hat{I}_k$ itself and the unit matrix,
times the energy $E$.

Then it is convenient to write the effective potential matrix as
\begin{eqnarray*}
E\{\hat{I}_k-1\}=E\{\hat{I}'_k-1\}+E\hat{I}''_k
\end{eqnarray*}
where the first matrix in the r.h.s. is proportional to the unit
matrix,
\begin{eqnarray}
E\{\hat{I}'_k-1\}=\left(\begin{array}{cc}
E{\lambda_1(k)+\lambda_2(k)-2\over2} & 0  \vspace{0.3cm} \\
0 & E{\lambda_1(k)+\lambda_2(k)-2\over2}
\end{array}\right).
\end{eqnarray}
This matrix does not couple $l$-channels and is influencing
elastic scattering phases only. The second matrix,
\begin{eqnarray}
E\hat{I}''_k=\left(\begin{array}{cc}
E{\lambda_1(k)-\lambda_2(k)\over2}\cos2\phi(k) & -
E{\lambda_1(k)-\lambda_2(k)\over2}\sin2\phi(k) \vspace{0.3cm}\\
-E{\lambda_1(k)-\lambda_2(k)\over2}\sin2\phi(k) &
-E{\lambda_1(k)-\lambda_2(k)\over2}\cos2\phi(k)
\end{array}\right)
\end{eqnarray}
is affecting the parameters of inelastic scattering.

These considerations are easily generalized to the case of
$^6$He$+^6$He, where there are five channels at any given $k$. In
particular, the elastic part of the effective potential matrix
becomes
\begin{eqnarray}
\label{v1} E\{\hat{I}'_k-1\}=E\cdot \sum_{i=1}^5
{\lambda_i(k)-1\over5} \left(\begin{array}{ccccc}
1 & 0 & 0 & 0 & 0  \\
0 & 1 & 0 & 0 & 0  \\
0 & 0 & 1 & 0 & 0  \\
0 & 0 & 0 & 1 & 0  \\
0 & 0 & 0 & 0 & 1
\end{array}\right).
\end{eqnarray}
It is clear (cf. Fig.~1) that
\begin{equation}
\label{effective}
 0 \le \Lambda_{\mbox{eff}} \le 1 , \,\,\,\,\,
\Lambda_{\mbox{eff}} \equiv \sum_{i=1}^5 {1-\lambda_i(k)\over5}
\end{equation}
If the energy $E$ is positive, the part (\ref{v1}) of the
effective potential is repulsive. Fig.~1 shows that the intensity
of the repulsion increases not only with the number of quanta $2k$
decreasing, which is expected, but also with the energy $E$
increasing, which is not. But perhaps this explains why the
elastic scattering phase infinitely increases with the energy.
Note that at any positive energy the elastic scattering in every
channel is over-the-barrier: the top of the barrier is $0.78E$ at
$k=2$.

\begin{figure}
\label{fig-1}
\vspace{-0.3cm}
\includegraphics[width=8cm]{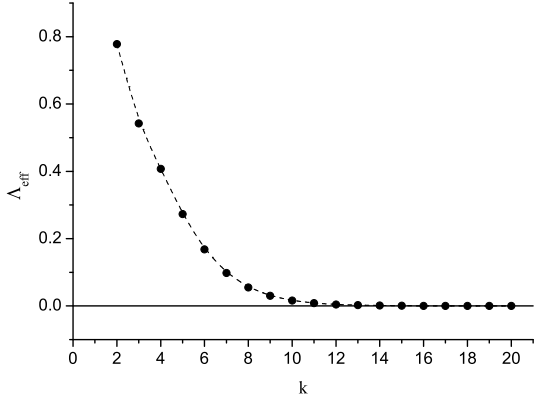}
\vspace{-0.5cm} \caption{Elastic part of the effective potential.
See Eq.(\ref{effective}) for the definition of
$\Lambda_{\mbox{eff}}$}
\end{figure}

The second, inelastic term of the effective potential depends on
the following combinations of the eigenvalues,
\begin{eqnarray*}
{4\over5}\left(\lambda_1(k)-{\lambda_2(k)+\lambda_3(k)+\lambda_4(k)+\lambda_5(k)\over4}\right),
\\ {3\over4}\left(\lambda_2(k)-{\lambda_3(k)+\lambda_4(k)+\lambda_5(k)\over3}\right),
\\
{2\over3}\left(\lambda_3(k)-{\lambda_4(k)+\lambda_5(k)\over2}\right),
~~{1\over2}\left(\lambda_4(k)-\lambda_5(k)\right).
\end{eqnarray*}
Each of these expressions vanishes at large $k$ and reaches its
maximum at the minimally allowed value of $k$.

In summary, the operator of antisymmetrization does not couple
$SU(3)$ channels. In the $l$-basis representation, the coupling of
channels via this operator decreases exponentially with $k$
increasing. Although the effective potential is a short-range one,
its range is several times more than the oscillator length $r_0$.
Its intensity is proportional to the energy of the continuous
states. In addition, it influences the inelasticity coefficients.
Therefore, the Pauli exclusion principle leads not only to the
repulsion of clusters at small distances, which has been
repeatedly discussed in the literature, but also to inelastic
scattering with excitation of clusters.

\section{Hamiltonian of the system $^6$He+$^6$He}

\subsection{Hamiltonian and the decay thresholds}

The Hamiltonian $\hat{H}$ of $^6$He+$^6$He is written as
\begin{eqnarray}
\hat{H}=\hat{h}_1+\hat{h}_2+\hat{T}+\hat{V}_{int},
\end{eqnarray}
where $\hat{h}_{i}$ is the Hamiltonian of the $i$th cluster,
$\hat{T}$ is operator of the kinetic energy of the relative motion
in the c.o.m. frame, and $\hat{V}_{int}$ is the interaction
between the clusters.

Based on the experimental evidence, we assume that the ground
state of $^6$He is an $s$-wave with the energy $e$, and there is a
resonance state\footnote{Its experimental width is only about 113
keV, so we treat it as a bound state here.} with $l=2$ and the
energy $e+\epsilon$. The energy $E$ of the $^{12}$Be is counted
from the threshold of its decay into two $^6$He nuclei in their
ground state. This decay channel is described by the $l$-basis
functions $\Phi_k^{(0,0,0)}$. Another threshold, that of the decay
with an excitation of one of the $^6$He nuclei to its $l=2$ state,
is located $\epsilon=1.8$ MeV above. This new open channel is
described by the functions
$${1\over\sqrt{2}}\left(\Phi_k^{(2,0,2)}+\Phi_k^{(0,2,2)}\right).$$
Finally, one more threshold, at $E=2\epsilon=3.6$ MeV, of the
decay with both fragments in their $l=2$ state. Above it, all five
channels are open.

Action of the operator $\hat{h}_1+\hat{h}_2$ to the $l$-basis
functions does not depend on the number of quanta $2k$,
\begin{eqnarray}
\label{V-me}
\left(\hat{h}_1+\hat{h}_2\right)\Phi_k^{(0,0,0)}&=&2e\Phi_k^{(0,0,0)};
\nonumber \\ \left(\hat{h}_1+\hat{h}_2\right)
{1\over\sqrt{2}}\left(\Phi_k^{(2,0,2)}+\Phi_k^{(0,2,2)}\right)&=&(2e+\epsilon)
{1\over\sqrt{2}}\left(\Phi_k^{(2,0,2)}+\Phi_k^{(0,2,2)}\right); \nonumber \\
\left(\hat{h}_1+\hat{h}_2\right)\Phi_k^{(2,2,l)}&=&
(2e+2\epsilon)\Phi_k^{(2,2,l)},~~l=0,2,4.
\end{eqnarray}
The threshold energies are taken into account by introduction of
the operator
$$\hat{V}=\hat{h}_1+\hat{h}_2-2e,$$
matrix elements of which in the $l$-basis are defined by
Eqs.(\ref{V-me}), and those in the $SU(3)$ basis are obtained
using the known relation between the two bases.

\subsection{Kinetic energy and the equations of free motion}

In the $l$-basis, the matrix elements of $\hat{T}$ are well-known,
\begin{eqnarray}
\label{kin:fiz}
\ME<l_1,l_2,l,2k+2|\hat{T}|l_1,l_2,l,2k>&=&-{1\over4}\sqrt{(2k-l+2)(2k+l+3)},
\nonumber \\
\ME<l_1,l_2,l,2k|\hat{T}|l_1,l_2,l,2k>&=&{1\over2}\left(2k+{3\over2}\right).
\end{eqnarray}
In the $SU(3)$ basis, the matrix elements are found using either
Eqs.(\ref{kin:fiz}) and the relations (\ref{unit}), or the
Fock--Bargmann map of the kinetic energy operator\cite{Fewbody},
\begin{eqnarray}
\label{kin1}
 \hat{T}_{\vec R} =-{1\over4}\left({\vec
R}^2-2({\vec R \cdot \vec \nabla_{\vec R} })-3+{\vec \nabla}_{\vec
R}^2\right) .
\end{eqnarray}


Consider now the expansion of the wave function $\Psi_{L=0}$ in
the $SU(3)$ basis,
\begin{eqnarray*}
\Psi_{L=0}=\sum_{k=2}^\infty C^{(2k-2,0)}_k\Psi^{(2k-2,0)}+
\sum_{k=3}^\infty C^{(2k-4,4)}_k\Psi^{(2k-4,4)}+
\end{eqnarray*}
\begin{eqnarray}
+\sum_{k=3}^\infty C^{(2k,2)1}_k\Psi^{(2k,2)1}+ \sum_{k=4}^\infty
C^{(2k,2)2}_k\Psi^{(2k,2)2}+ \sum_{k=5}^\infty
C^{(2k+4,0)}_k\Psi^{(2k+4,0)}.
\end{eqnarray}
The expansion coefficients  $C^{j}_k\equiv
C^{(\lambda,\mu)\nu}_k,~j=\overline{1,5}$ satisfy the set of
linear algebraic homogeneous equations,
\begin{eqnarray}
\label{u8} \sum_{k',j'}\ME<k,j|{\hat H}|k',j'>
C^{j'}_{k'}-E\Lambda^j_{k}\,C^{j}_k= 0,~~{\hat H}={\hat T}+{\hat
V} + {\hat V}_{int}.
\end{eqnarray}

In the limit $k\rightarrow\infty$ (effectively, large distances
between the clusters), the interaction between the clusters is
negligible, and the set (\ref{u8}) with ${\hat V}_{int}=0$ is
defining the asymptotic behavior of the coefficients $C^j_k$.

And again, utilizing the relations (\ref{unit}) one can find the
asymptotic values of the expansion coefficients $C^{j}_k$ knowing
those in the $l$-basis, $C_k^{(l_1,l_2,l)}$. At small values of
$k$, the set (\ref{u8}) has a cumbersome form in either basis. In
the domain of large $k$, however, it decouples into five
independent equations in the $l$-basis,
\begin{eqnarray*}
-{1\over4}\sqrt{(2k-l+2)(2k+l+3)}C^{(l_1,l_2,l)}_{k+1}
-{1\over4}\sqrt{(2k-l)(2k+l+1)}C^{(l_1,l_2,l)}_{k-1}+
\end{eqnarray*}
\begin{eqnarray}
+\left\{{1\over2}\left(2k+{3\over2}\right)-(E-E_i)\right\}C^{(l_1,l_2,l)}_k=0,
\end{eqnarray}
each of which having a limiting ($k\rightarrow\infty$) form of the
Bessel equation, whereas in the $SU(3)$ basis the equations remain
coupled\cite{Fewbody}.

The asymptotic form of the expansion coefficients in the $l$-basis
can be conveniently written in terms of the Hankel functions
$H^{\pm}_{l+1/2}$. If the incoming wave is in the channel
($l_1,~l_2,~l$), the expansion coefficients in this channel
satisfy the asymptotic relation
\begin{eqnarray}
\label{asymp} C^{(l_1,l_2,l)}_k\equiv C_i^i(\kappa_i
\sqrt{4k+3})=H^-_{l+1/2}(\kappa_i \sqrt{4k+3})+
S_{ii}H^+_{l+1/2}(\kappa_i \sqrt{4k+3}),
\end{eqnarray}
where $i=1, \dots, 5$,
$$H^{\pm}_{l+1/2}(\kappa_i \sqrt{4k+3})=J_{l+1/2}(\kappa_i \sqrt{4k+3})\pm
iN_{l+1/2}(\kappa_i \sqrt{4k+3}),~~\kappa_i^2=2(E-E_i),$$
$E_i$ is the threshold energy of the $i$th channel
$(E_i=0,~\epsilon,~2\epsilon)$, $S_{ij}$ are the scattering matrix
elements. (We here put the nucleon mass and the Planck's constant
equal to 1 for the sake of brevity.)

The asymptotic behavior of the expansion coefficients in the exit
channels ($\tilde{l}_1,~\tilde{l}_2,~\tilde{l}$) is defined by
\begin{eqnarray}
\label{asymp2}
 C^{(\tilde{l}_1,\tilde{l}_2,\tilde{l})}_k\equiv C_j^i(\kappa_j
\sqrt{4k+3})=S_{ij}H^+_{l+1/2}(\kappa_j
\sqrt{4k+3}),~~i=\overline{1,5},~j\neq i.
\end{eqnarray}
Since the coefficients $C^{j}_k$ and $C_k^{(l_1,l_2,l)}$ are
related via the same unitary matrix (\ref{unit}) as the basis
functions themselves, it is easy to find the asymptotic values of
$C^{j}_k$ with Eqs.(\ref{asymp}), (\ref{asymp2}), (\ref{unit}) and
then solve the set (\ref{u8}).

\section{Ground state of $^{12}$Be}

As we stated earlier, we restricted our study to the $L=0$ states
of $^{12}$Be. Another restriction comes from the use of the
approximation of the zero-range nuclear force\cite{Bete}; we
assume that the interaction can be reproduced by just two diagonal
matrix elements in the $SU(3)$ representation $(2k-2,0)$, i.e.
$$
\ME<(2k-2,0)|U|(2k'-2,0)> = \left\{ \begin{array}{cc} - 44.2
\mbox{ MeV} & \mbox{ if } k=k'=2 \\ - 28.7 \mbox{ MeV} & \mbox{ if
} k=k'=3 \\0  & \mbox{ otherwise }   \end{array} \right.
$$
These values were fitted to the experimental values of the r.m.s.
radius of $^{12}$Be ($2.59\pm0.06$ fm\cite{Tan}) in its ground
state, and the $^6$He$+^6$He decay threshold ($10.11$
MeV\cite{Freer2}). These were the only numerical parameters in our
model; the oscillator length was fixed to $1.37$ fm.

\begin{figure}
\label{fig-2} \vspace{-0.3cm}
\includegraphics[width=10cm]{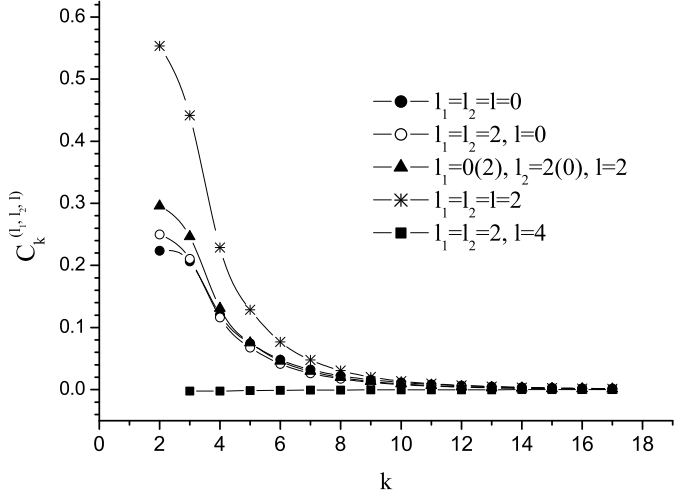}
\vspace{-0.5cm} \caption{Ground state of $^{12}$Be: coefficients
of the w.f. expansion in the $l$-basis}
\end{figure}

We were able to extract some information on the structure of the
ground state of $^{12}$Be from the coefficients of the expansion
of its wave function over the Pauli-allowed basis states (Fig.~2).
Firstly, this nucleus appears to be softer than it would follow
the results of the shell-model calculations. The standard
shell-model configuration corresponds to the allowed state with
the minimal number of quanta. The weight of this component in the
g.s. wave function is only 54\%. The considerable contribution of
other states even at relatively large $k$ indicates the
diffuseness of the nuclear surface and shows the correct
asymptotic behavior of the wave function in each of the five
closed channels.

Imitating the dependence of the wave function in the coordinate
representation on the inter-cluster distance, the expansion
coefficients fall exponentially with the number of quanta
increasing. Asymptotically, in the $i$th $l$-channel,
\begin{eqnarray}
C^i_0(k)\sim A^i_0\sqrt{2}\,{\exp\{-\sqrt{2(E_i-E_0)}\sqrt{4k+3}\}
\over{\sqrt[4]{4k+3}}},
\end{eqnarray}
where $E_i$ is the threshold energy for this channel, $E_0$ is the
g.s. energy of $^{12}$Be. The factor $A^i_0$ is usually called
"asymptotic normalization coefficient" (ANC)\footnote{There is a
limiting expression for the normalized radial three-dimensional
oscillator functions,
$$R_{nl}(r)r^{3/2}\sim \sqrt{2}\delta(r-r_0\sqrt{4n+2l+3}).$$
In the short-range potential, the radial part of the wave function
of a bound state behaves like
\begin{eqnarray}
\Psi^i_{l}(r)\sim A^i_l{\exp(-\alpha_ir)\over r},
\end{eqnarray}
if $r$ is much greater than the range of the potential. Here
$\alpha_i=\sqrt{2E_i}$, $A^i_l$ is the asymptotic normalization
coefficient. The expansion coefficients of the function
$\Psi^i_{l}(r)$ in the harmonic-oscillator basis are defined as
\begin{eqnarray}
C^i_{nl}=\int\Psi^i_{l}(r)R_{nl}(r)r^2dr.
\end{eqnarray}
Therefore, if the number of quanta is $n\gg1$, they are expressed
in terms of the asymptotic normalization coefficients,
\begin{eqnarray}
C^i_{nl}=A^i_l\sqrt{2}\,{\exp(-\alpha_i r_0\sqrt{4n+2l+3})\over
\sqrt{r_0}\sqrt[4]{4n+2l+3}}.
\end{eqnarray} } \cite{Bloch}.

The function of the channel ($l_1=l_2=l=2$) is characterized by
the largest value of ANC, 43.49 fm$^{-1/2}$. In the channels
($l_1=l_2=l=0$), ($l_1=l_2=2,~l=0$) and ($l_1=2 (0), l_2=0
(2),~l=2$) the values of ANC are, respectively, 12.99 fm$^{-1/2}$,
25.61 fm$^{-1/2}$ and 19.1 fm$^{-1/2}$. The smallest ANC (0.8
fm$^{-1/2}$) is in the channel ($l_1=l_2=2,~l=4$).

Due to the large value of its ANC, the channel ($l_1=l_2=l=2$) is
dominating in the g.s. function, with the weight 58\%, despite the
fact that its threshold is higher than those of ($l_1=l_2=l=0$)
and ($l_1=2 (0), l_2=0 (2),~l=2$). The contributions of the latter
channels are 12\% and 17\%, respectively. Those of the
($l_1=l_2=2,~l=0$) and ($l_1=l_2=2,~l=4$) channels are 13\% and
less than $10^{-5}\%$, respectively.

We also studied the expansion of the shell-model g.s. wave
function of $^{12}$Be (the $(\lambda,\mu)=(2,0)$ function of the
$SU(3)$ basis) in the $l$-basis. It is dominated by the
($l_1=l_2=l=2$) component -- 60\%, followed by ($l_1=2 (0),~l_2=0
(2),~l=2$) -- $17.5 \%$, ($l_1=l_2=2,~l=0$) -- $12.5\%$,
($l_1=l_2=l=0$) -- 10\%.

\section{Continuum part of the spectrum}

In the approximation used here, the continuum part of the
$^{12}$Be spectrum begins over the threshold of the decay into two
$^6$He nuclei in their ground states. We set the energy of this
threshold to zero. In the interval of energies $0<E<\epsilon$, the
elastic scattering of two $^6$He nuclei is the only open channel,
and all the information about this process is in the only
$S$-matrix element $S_{11}=\exp(2i\delta_{11})$, i.e. in the
scattering phase $\delta_{11}(E)$ (Fig.~3). The approximation and
the limited, although infinite, basis provides for the existence
of the sole bound state, therefore the phase $\delta_{11}(0)$ is
set to $\pi$. In the vicinity of the threshold, the derivative of
the phase with respect to the energy is inversely proportional to
the square root of the energy, and the scattering length
$$a=-\lim{\delta_{11}(\kappa)\over \kappa}=6.62 \mbox{fm},~~\kappa=\sqrt{2E}
\rightarrow 0.$$ The phase steadily decreases with the energy
increasing. Only a considerable rise in the intensity of the
attractive potential can slow down this fall or force an increase.

\begin{figure}
\label{fig-3}
\centering
\begin{minipage}[c]{.55\textwidth}
\centering \vspace{-0.3cm}
\includegraphics[width=8cm]{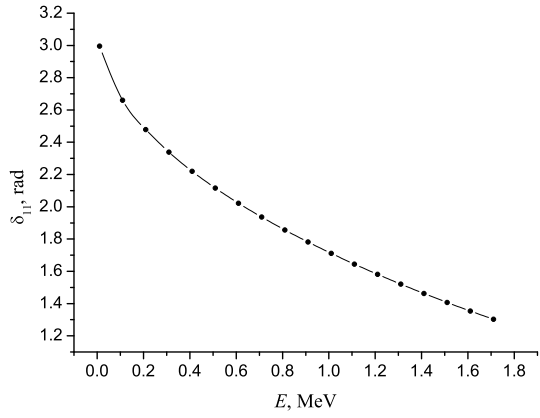}
\end{minipage}
\begin{minipage}[c]{.40\textwidth}
\vspace{-0.3cm} \centering \caption{Subthreshold elastic
scattering in the channel ($l_1=l_2=l=0$).}
\end{minipage}
\end{figure}
\begin{figure}
\label{fig-4} \centering
\begin{minipage}[c]{.55\textwidth}
\centering \vspace{-0.3cm}
\includegraphics[width=7.5cm]{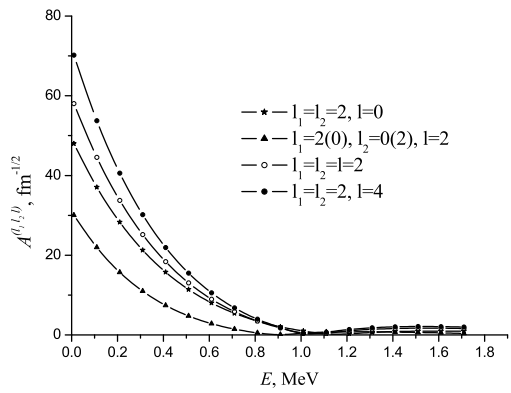}
\end{minipage}
\hfill
\begin{minipage}[r]{.40\textwidth}\vspace{-0.3cm}
\caption{Asymptotic normalization coefficients (ANCs) in various
exit channels. The entrance channel is ($l_1=l_2=l=0$).}
\end{minipage}
\end{figure}

As long as the energy is less than $\epsilon$, the behavior of the
wave function in the closed channels is reproduced by the energy
dependence of the four ANCs shown at Fig.~4. Their values are
several times larger than those of the g.s. function. They reach
their maxima just over the threshold (the largest ANC there, about
70 fm$^{-1/2}$, belongs to the ($l_1=l_2=2, l=4$) channel), and
then they steadily fall, until the next threshold is reached.

\begin{figure}[h]
\label{fig-5} \centering
\begin{minipage}[c]{.45\textwidth}\vspace{-0.2cm}
\hspace{-0.3cm}
\includegraphics[width=7cm]{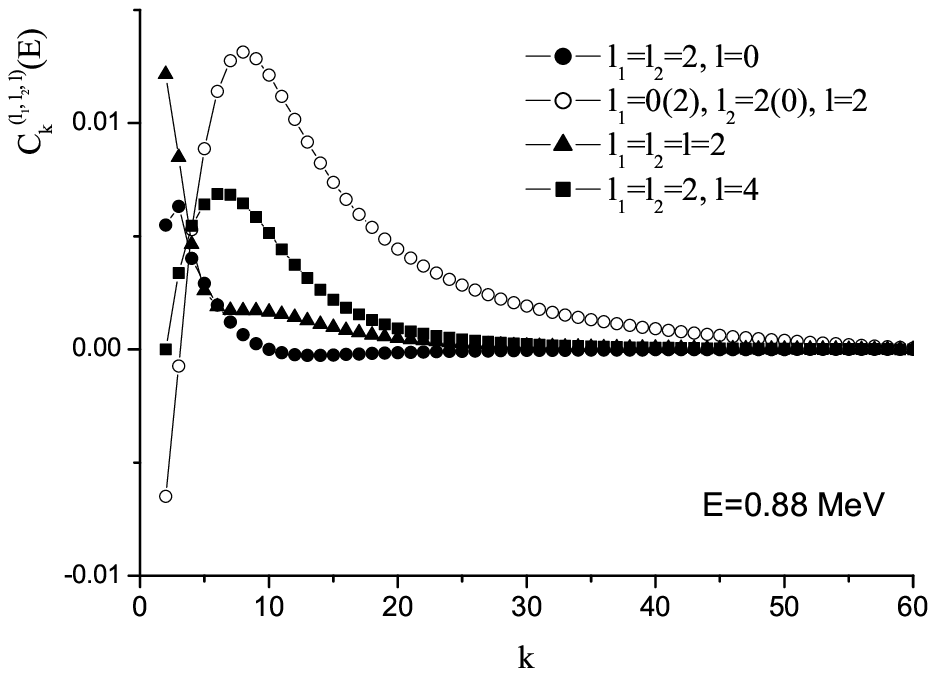}
\end{minipage}
\hfill
\begin{minipage}[c]{.45\textwidth}\vspace{-0.3cm}
\hspace{-0.8cm}
\includegraphics[width=7cm]{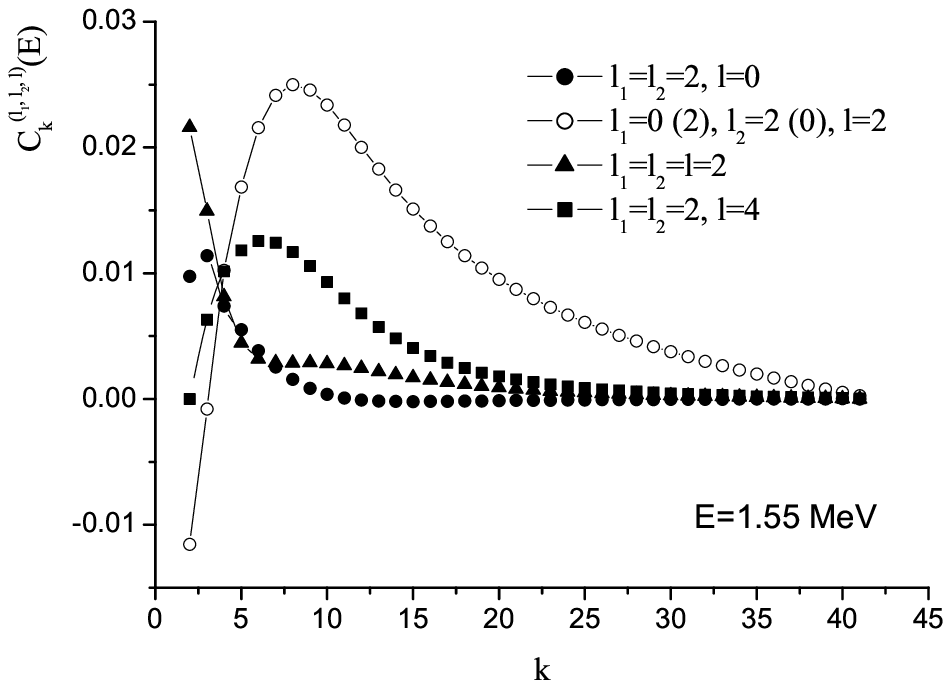}
\end{minipage}
\caption{Coefficients $C^{(l_1,l_2,l)}_k(E)$ of the expansion of
two continuum states (at $E=0.88$ MeV, left, and at $E=1.55$ MeV,
right) in the $l$-basis}
\end{figure}

When the energy reaches $\epsilon=1.8$ MeV, the channel ($l_1=2
(0), l_2=0 (2),~l=2$) opens. Just below this threshold the wave
function of this channel is expected to fall slower than those in
other channels. Indeed (Fig.~5), the state ($l_1=2 (0), l_2=0
(2),~l=2$) dominates and has the longest tail. This dependence
becomes more pronounced as the energy approaches the threshold.

Between the second and third thresholds the $S$-matrix has a
standard two-channel form,
\begin{eqnarray}
\label{scat} S=\left|\begin{array}{cc}
S_{11} & S_{12} \\
S_{21} & S_{22}
\end{array}
\right|
\end{eqnarray}
\begin{eqnarray*}
S_{11}=\eta\exp(2i\delta_{11}),~~~S_{12}=i\sqrt{1-\eta^2}\exp(i\delta_{11}+
i\delta_{22}),
\end{eqnarray*}
\begin{eqnarray*}
S_{21}=i\sqrt{1-\eta^2}\exp(i\delta_{11}+i\delta_{22}),~~~S_{22}=
\eta\exp(2i\delta_{22}).
\end{eqnarray*}
In this energy region, $\epsilon<E<2\epsilon$, the three ANCs are
an order of magnitude smaller than before. They depend not only on
the energy, but also on which of the two open channels is the
entrance one. Both dependences are illustrated at Fig.~6.

\begin{figure}[h]
\label{fig-6} \centering
\begin{minipage}[c]{.45\textwidth}
\vspace{-0.4cm}
\includegraphics[width=6.5cm]{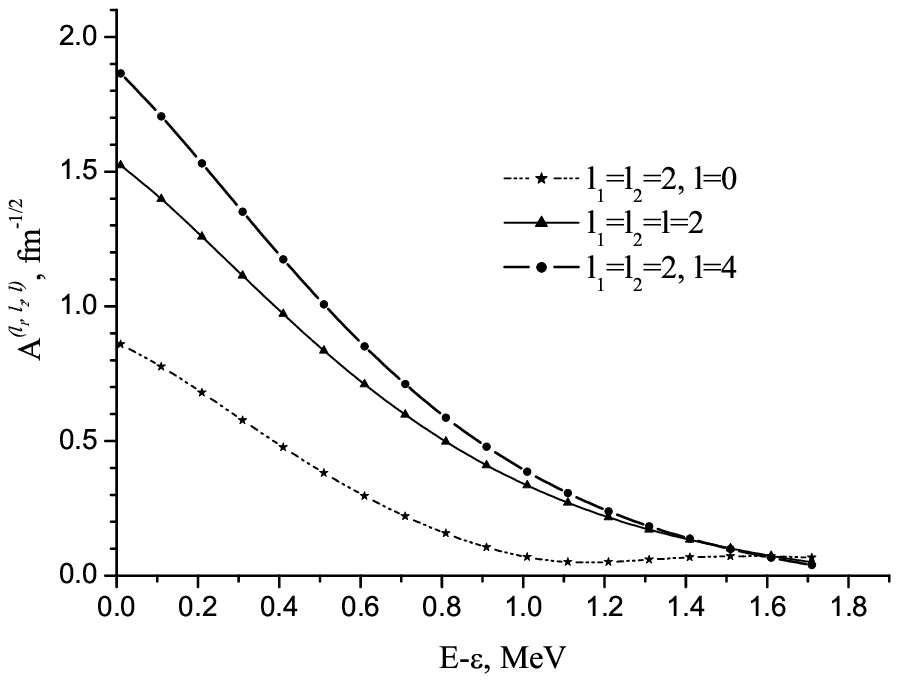}
\end{minipage}
\hfill
\begin{minipage}[c]{.50\textwidth}
\vspace{-0.4cm}\hspace{-0.3cm}
\includegraphics[width=7.3cm]{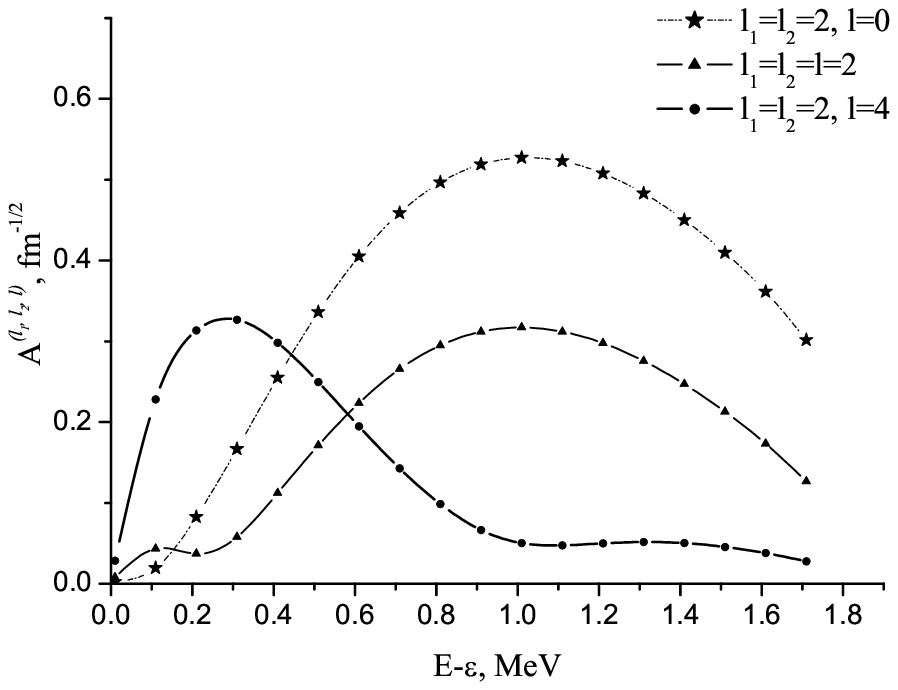}
\end{minipage}
\caption{Asymptotic normalization coefficients (ANCs) in various
exit channels in the energy domain between two thresholds. The
entrance channel are ($l_1=l_2=l=0$) (left) and ($l_1=2 (0), l_2=0
(2),~l=2$) (right)}
\end{figure}

In the left panel of Fig.~6, the entrance channel is
($l_1=l_2=l=0$). The ANCs decrease monotonically with increasing
energy, reaching their maximum values when $E-\epsilon$ is close
to zero. When $E-\epsilon$ is small, the absolute values of the
ANC in the channels ($l_1=l_2=2, l=4$) and ($l_1=l_2=l=2$) are
almost twice larger than that in the channel ($l_1=l_2=2, l=0$).

If the entrance channel is ($l_1=2 (0), l_2=0 (2),~l=2$), the
energy dependence is more interesting (Fig.~6, right panel). The
ANCs in the channels ($l_1=l_2=2,~l=0$) and ($l_1=l_2=l=2$) have
broad peaks at $E-\epsilon=1$ MeV, reaching 0.5 fm$^{-1/2}$ and
0.3 fm$^{-1/2}$, respectively, and then fall. In the
($l_1=l_2=2,~l=4$) channel, the ANC has a more pronounced maximum
at $E-\epsilon=0.3$ MeV and than falls to almost zero. Such a
behavior may indicate the existence of a resonance in this
channel.

Above the third threshold, where $E>2\epsilon$, all the five
channels are open, and the $S$-matrix is $5\times5$. Having
computed its elements, we found the effective cross-sections of
elastic and inelastic scattering for energies up to 30 MeV.

The cross-section $\sigma_{11}$ of the elastic scattering in the
channel ($l_1=l_2=l=0$) is shown at Fig.~7. It decreases smoothly
from 5.1 b at $E=0.1$ MeV to 14 mb at $E=5$ MeV. There are no
visible peculiarities at the thresholds $E=\epsilon$ and
$E=2\epsilon$. However it is known that the existence of threshold
reactions may be exhibited in a characteristic dependence of the
elastic cross-section on the energy around the threshold (the
Wigner--Baz' effect\cite{Davydov}).

\begin{figure}
\label{fig-7} \centering
\begin{minipage}[c]{.55\textwidth}
\centering
\includegraphics[width=8cm]{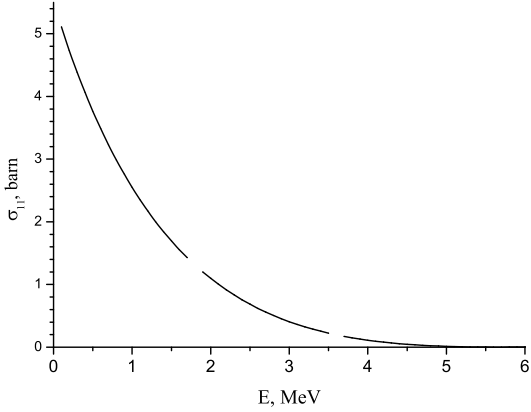}
\end{minipage}
\hfill
\begin{minipage}[r]{.30\textwidth}
\caption{Cross-section of the elastic scattering in the channel
($l_1=l_2=l=0$). The breaks in the line indicate the decay
thresholds}
\end{minipage}
\end{figure}

Consider the two-channel scattering matrix, Eq.(\ref{scat}). If,
at $E=\epsilon$, there opens a channel with the relative orbital
momenta $l$, then at small positive values of
$E-\epsilon=\kappa_1^2/2$ the dependence of the inelasticity
coefficient $\eta$ on the wave number $\kappa_1$ is defined by
$$\eta\sim 1-b\kappa_1^{2l+1}-...,~~b>0.$$
In this case, the elastic scattering in the energy region over the
threshold $E=\epsilon$ should take the form
\begin{eqnarray*}
\sigma_{11}= {4\pi\over
\kappa^2}\sin^2\delta_{11}(1-a\kappa_1^{2l+1}-...).
\end{eqnarray*}
Under this threshold, the $S$-matrix is reduced to a single value,
and the cross-section follows the law
\begin{eqnarray}
\sigma_{11}={4\pi\over \kappa^2}\sin^2\delta_{11}.
\end{eqnarray}
Therefore, at the threshold one may expect to see a change in the
monotonic behavior. Evidently, this effect will be significant
only if $l=0$. However, in the case considered here, $l=2$. So in
order to study the Wigner--Baz' effect we have checked the third
threshold, $E=2\epsilon$, because one of the channels which open
there has $l=0$. Still, there are no peculiarities in the
cross-section. The explanation is following.

It appears that, in the expansion of the inelasticity coefficients
$\eta_{1l}$ in the domain of small positive values of
$E-2\epsilon=\kappa_2^2/2$, the first non-vanishing terms that
define the behavior of the cross-sections are proportional to
$b_l\kappa_2^{2l+3}$ rather than $b_l\kappa_2^{2l+1}$. Therefore,
the Wigner-Baz' effect will be less pronounced, and the following
behavior is expected above the threshold $E=2\epsilon$,
\begin{eqnarray}
\sigma_{1l}={\pi\over
k^2}~b_l\left\{2(E-2\epsilon)\right\}^{l+3/2}, ~l=0,~2,~4.
\end{eqnarray}
\begin{figure}
\label{fig-8} \centering
\begin{minipage}[c]{.45\textwidth}
\vspace{-0.3cm}\hspace{-0.4cm}
\includegraphics[width=7.2cm]{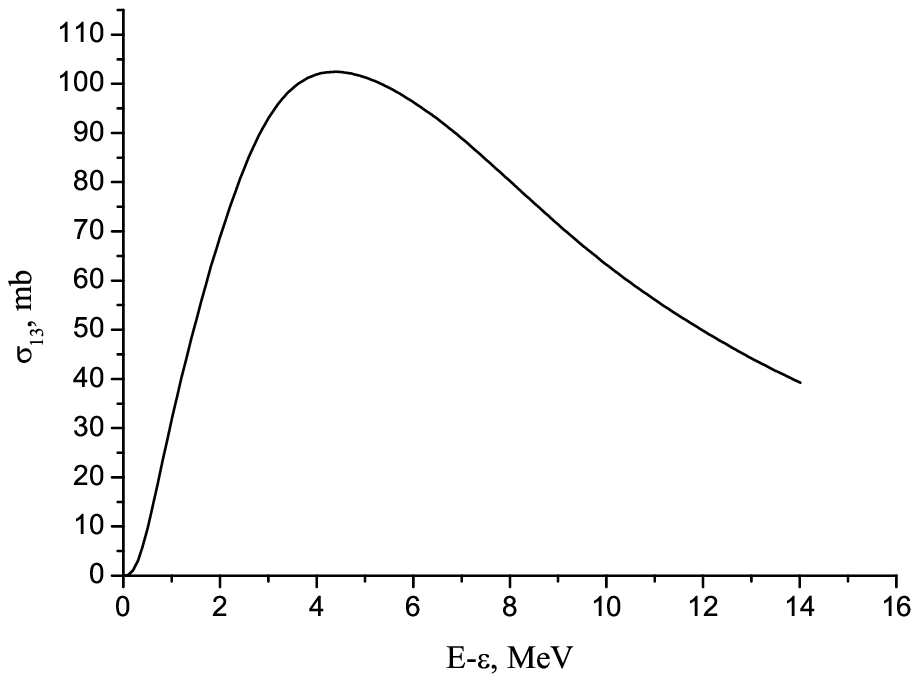}
\end{minipage}
\hfill
\begin{minipage}[c]{.50\textwidth}
\vspace{-0.3cm}\hspace{-0.4cm}
\includegraphics[width=7.2cm]{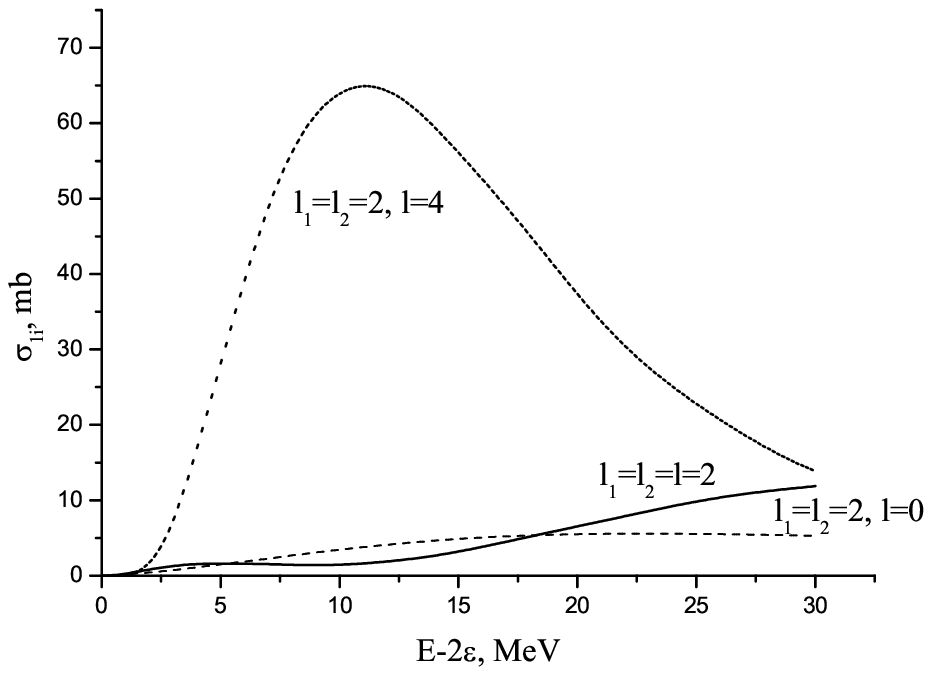}
\end{minipage}
\vspace{-0.3cm} \caption{Cross-sections of the inelastic
scattering into the channels ($l_1=2 (0), l_2=0 (2),~l=2$) (left),
and ($l_1=l_2=2, l=0,~2,~4$) (right). The entrance channel is
($l_1=l_2=l=0$)}
\end{figure}
In Fig.~8, inelastic cross-sections with the exit channels ($l_1=2
(0), l_2=0 (2),~l=2$) $(\sigma_{13})$ and ($l_1=l_2=2, l=0,~2,~4$)
are compared. The behavior of the cross-sections does not indicate
the existence of a resonance over the threshold
$^{12}$Be$\rightarrow^6$He+$^6$He up to several dozen MeV. The
cross-section $\sigma_{13}$ has a broad maximum at 4.4 MeV over
the threshold $E=\epsilon$, reaching 102 mb, and falls to 39 mb at
$E-\epsilon=14$ MeV. Over the threshold $E=2\epsilon$, the exit
channel ($l_1=l_2=2,~l=4$) dominates. The corresponding
cross-section reaches 65 mb at $E-2\epsilon\sim11$ MeV, while the
other cross-sections do not exceed 12 mb.

\section{Conclusions}

Taking a system of two $^6$He clusters in $^{12}$Be as an example,
we have shown that in the Fock--Bargmann space the eigenvalues of
the norm kernel can be found by integrating its products with its
eigenfunctions analytically. The latter are functions of the
$SU(3)$ basis and can be constructed {\it a priori}, if there is
no $SU(3)$-degeneracy. If there is one, the eigenfunctions are
found as solutions of an integral equation with a degenerate
kernel. When solving this equation, one uses standard algebraic
procedures. An important detail is that the eigenvalues of this
integral equation have a finite limiting point, where the
eigenfunctions are defined with an accuracy up to a unitary
transform. On the other hand, at any finite number of oscillator
quanta $2k$ these functions are unique. At small $k$,
eigenfunctions of a degenerate $SU(3)$ representation are related
to their asymptotic counterparts (i.e., eigenfunctions at
$k\rightarrow\infty$) via a rotation matrix. The angle of the
rotation varies with $k$, changing the structure of the degenerate
states.

The eigenvalues of the norm kernel limit to unity with $k$
increasing, with exponentially small corrections. Nevertheless,
these corrections are important for the unique determination of
the asymptotic eigenfunctions.

With the number of quanta increasing, the number of Pauli-allowed
states grows from one (at $n=2k=4$) to five ($n \ge 10$).

Taking into account the Pauli principle leads to an effective
potential related to the antisymmetrization. In the representation
of the angular momentum-coupled $l$-basis, this potential consists
of two terms. One of the terms determines the elastic scattering
cross-section; it is a repulsion, the intensity of which is
proportional to the energy of the continuum states and increases
with the number of quanta decreasing. The scattering occurs over
the barrier at any positive energy in all channels. The range of
the repulsion is several times larger than the value of the
oscillator radius, as deduced from the large scattering length and
the elastic cross-section reaching several barn.

The second term influences the inelastic cross-sections. In the
$l$-basis representation, the channels are coupled by the
antisymmetrizer; the coupling falls exponentially with $k$
increasing, and the inelastic cross-sections are an order of
magnitude smaller than the elastic one (several dozen mbarn).

In the representation of the $SU(3)$ basis, meanwhile, the
channels are coupled not by the antisymmetrizer, but by the
kinetic energy operator. With $k$ increasing, this coupling
decreases more slowly, as $1/k$. The unitary transformation from
the $SU(3)$ basis to the $l$-basis decouples the asymptotic
equations of AVRGM, thus allowing to solve these equations in
either basis.

Due to the Pauli principle, the wave functions of both the ground
state and the continuum states of $^{12}$Be are distributed over
several $l$-channels. The contribution of each channel is
determined by the value of the normalization coefficient (related
to the amplitude of the wave function in a closed channel) and by
the proximity of the channel threshold. Thus the ground state wave
function is dominated by the ($l_1=l_2=l=2$) channel due to its
large normalization coefficient. It is also important to note the
softness of the $^{12}$Be nucleus in comparison with the
shell-model predictions.

The behavior of the normalization coefficients in the energy
domain $\epsilon<E<2\epsilon$ of the continuum depends not only on
the energy, but also on which of the two open channels is the
entrance channel. A pronounced peak is observed in the dependence
of the normalization coefficient on the energy in the channel
($l_1=l_2=2,~l=4$), provided that the entry channel is
$^6$He$+^6$He$^*$. This may signal the existence of a resonance in
this channel.

The behavior of scattering phases does not indicate that there are
resonances in any of the channels.

\bigskip

\appendix

\section{$SU(3)$ invariants $F_{(\lambda,\mu)\nu}$}

The norm kernel of $^6$He+$^6$He can be expanded over
$SU(3)$-scalar blocks $F_{(\lambda,\mu)\nu}$ (cf. Eq.(\ref{s8})).
Explicit expressions (Eq.\ref{s4}) for these invariants are shown
below. A shorthand notation
$$
\Phi_{(\lambda', \mu')\nu'+\nu''} \equiv \Phi_{(\lambda',
\mu')\nu'} + \Phi_{(\lambda', \mu')\nu''}
$$
is used.

\begin{eqnarray*}
F_{(n+4,0)} &=& {1 \over 4n!} \left\{ \Phi_{(n+4,0)} - {4 \over
n+4} \,\Phi_{(n+2,1)1} - {2 \over n+4}\, \Phi_{(n+2,1)2+3}\right.
+
{2 \over (n+3)(n+4)} \,\Phi_{(n,2)1} \\
&&+  {n(n-1) \over (n+3)(n+4)} \,\Phi_{(n,2)2+3}
 + {4n(n-1) \over
(n+3)(n+4)} \,\Phi_{(n,2)4}  + {4n \over (n+3)(n+4)}
\,\Phi_{(n,2)5+6}
\\ && - {2n(n-1)(n-2) \over (n+2)(n+3)(n+4)}
\,\Phi_{(n-2,3)1+2} - {4n(n-1) \over (n+2)(n+3)(n+4)}
\,\Phi_{(n-2,3)3}
\\
&& + {n(n-1)(n-2)(n-3) \over (n+1)(n+2)(n+3)(n+4)}
\,\Phi_{(n-4,4)} + {4n \over (n+3)(n+4)} \,\Phi_{(n+1,0)} \\
&& - {4n \over (n+2)(n+3)(n+4)} \,\Phi_{(n-1,1)1} - {4n(n-1) \over
(n+2)(n+3)(n+4)} \,\Phi_{(n-1,1)2+3}
\\
&& + {4n(n-1)(n-2) \over (n+1)(n+2)(n+3)(n+4)} \,\Phi_{(n-3,2)}
\\ && \left. +
{2n(n-1) \over (n+1)(n+2)(n+3)(n+4)} \,\Phi_{(n-2,0)} \right\} \\
F_{(n+2,1)1} &=& {1 \over 4n!} \left\{ \Phi_{(n+2,1)1}
 -
{1 \over n+2} \,\Phi_{(n,2)1} - {n \over n+2} \,\Phi_{(n,2)5+6}+
{n(n-1) \over (n+1)(n+2)} \,\Phi_{(n-2,3)3} \right.
\\ &&
 - {n^2 \over (n+2)(n+4)} \,\Phi_{(n+1,0)} + {2n \over (n+1)(n+4)} \,\Phi_{(n-1,1)1}\\
&&   +{n^2(n-1) \over (n+1)(n+2)(n+4)} \,\Phi_{(n-1,1)2+3}-
{n(n-1)(n-2) \over (n+1)(n+2)(n+4)} \,\Phi_{(n-3,2)}
\\
&&  \left. -
{n(n-1) \over (n+1)(n+2)(n+4)} \,\Phi_{(n-2,0)} \right\}\\
F_{(n+2,1)2} &=& {1 \over 4(n+4)(n-1)!} \left\{  2 \,
\Phi_{(n+2,1)2+3} - \Phi_{(n+2,1)1}  +
{1 \over n+2} \,\Phi_{(n,2)1} \right. \\
&&-  {2(n-1) \over n+2} \,\Phi_{(n,2)2+3}
 - {8(n-1) \over n+2} \,\Phi_{(n,2)4}  + {n-4 \over n+2}
\,\Phi_{(n,2)5+6}
\\ && + {6(n-1)(n-2) \over (n+1)(n+2)}
\,\Phi_{(n-2,3)1+2} - {(n-1)(n-8) \over (n+1)(n+2)}
\,\Phi_{(n-2,3)3}
\\
&& - {4(n-1)(n-2)(n-3) \over n(n+1)(n+2)}
\,\Phi_{(n-4,4)} + {n-4 \over n+2} \,\Phi_{(n+1,0)} \\
&& - {2(n-2) \over (n+1)(n+2)} \,\Phi_{(n-1,1)1} - {(n-1)(n-8)
\over (n+1)(n+2)} \,\Phi_{(n-1,1)2+3}
\\ && \left. + {(n-1)(n-2)(n-12) \over n(n+1)(n+2)} \,\Phi_{(n-3,2)} +
{(n-1)(n-4) \over n(n+1)(n+2)} \,\Phi_{(n-2,0)} \right\} \\
F_{(n,2)1} &=& {1 \over 12 n!} \left\{ \Phi_{(n,2)1}
 -
{2n \over n+3} \,\Phi_{(n-1,1)1} +
{n(n-1) \over (n+2)(n+3)} \,\Phi_{(n-2,0)} \right\} \\
F_{(n,2)2} &=& {1 \over 8(n+2) (n-1)!} \left\{ 2 \Phi_{(n,2)5+6} -
\Phi_{(n,2)1}
 -
{4(n-1) \over n} \,\Phi_{(n-2,3)3}  -2 \,\Phi_{(n+1,0)} \right. \\
&& + {2(n-1)(n+2) \over n(n+3)} \,\Phi_{(n-1,1)1} + {4(n-1) \over
n(n+3)}
\,\Phi_{(n-1,1)2+3} \\
&&   \left. + {2(n-1)(n-2) \over n(n+3)} \,\Phi_{(n-3,2)}+
{n-1 \over n+3} \,\Phi_{(n-2,0)} \right\} \\
F_{(n,2)3} &=& {1 \over 4(n+2)(n+3) (n-2)!} \left\{ {1\over6} \,
\Phi_{(n,2)1} + \Phi_{(n,2)2+3} + 4 \,\Phi_{(n,2)4} -
\Phi_{(n,2)5+6} \right. \\ && - {6(n-2) \over n}
\,\Phi_{(n-2,3)1+2} + {2(n-3) \over n} \,\Phi_{(n-2,3)3}  +
{6(n-2)(n-3) \over n(n-1)} \,\Phi_{(n-4,4)}  \\ && -
\Phi_{(n+1,0)} - {n-6 \over 3n} \,\Phi_{(n-1,1)1}  +  {2(n-3)
\over n}
\,\Phi_{(n-1,1)2+3}   \\
&&   \left. - {3(n-2)(n-5) \over n(n-1)} \,\Phi_{(n-3,2)} +
{n^2-13n+24 \over 6n(n-1)} \,\Phi_{(n-2,0)} \right\} \\
F_{(n-2,3)1} &=& {n-1 \over 12(n+1)!} \left\{ 3 \Phi_{(n-2,3)3}  -
\Phi_{(n-1,1)1+2+3}
- {n-2 \over n+2} \,\Phi_{(n-3,2)}-
{n \over n+2} \,\Phi_{(n-2,0)} \right\} \\
F_{(n-2,3)2} &=& {(n-1)(n-2) \over 2(n+2)!} \left\{
\Phi_{(n-2,3)1+2} - {1\over2} \, \Phi_{(n-2,3)3} - {2(n-3) \over
n-2} \,\Phi_{(n-4,4)} \right.\\
&&  + {1 \over 6} \,\Phi_{(n-1,1)1}-{1 \over 2}
\,\Phi_{(n-1,1)2+3}    \left. - {n-2 \over 3(n+2)}
\,\Phi_{(n-3,2)}+
{n \over 3(n+2)} \,\Phi_{(n-2,0)} \right\} \\
F_{(n-4,4)} &=& {n-3 \over 4(n+1)!} \left\{ \Phi_{(n-4,4)} -
\Phi_{(n-3,2)}+
{1 \over 6} \,\Phi_{(n-2,0)} \right\} \\
F_{(n+1,0)} &=& {1 \over 2(n+4)(n-1)!} \left\{ \Phi_{(n+1,0)} -
{1\over n+1 } \, \Phi_{(n-1,1)1} -{n-1\over n+1 } \,
\Phi_{(n-1,1)2+3} \right. \\ && \left. +{(n-1)(n-2) \over n(n+1)}
\, \Phi_{(n-3,2)} +
{n-1 \over n(n+1)} \,\Phi_{(n-2,0)} \right\} \\
F_{(n-1,1)1} &=& {1 \over 6(n+3)(n-1)!} \left\{ \Phi_{(n-1,1)1} -
{n-1 \over (n+1)} \,\Phi_{(n-2,0)} \right\} \\
F_{(n-1,1)2} &=& {1 \over 6(n+1)(n+3)(n-2)!} \left\{
2\,\Phi_{(n-1,1)2+3}-\Phi_{(n-1,1)1} \right. \\ && \left. -
{4(n-2) \over n-1} \Phi_{(n-3,2)} +
{n-3 \over n-1} \,\Phi_{(n-2,0)} \right\} \\
F_{(n-3,2)} &=& {(n+1)(n-2) \over 4(n+2)!} \left\{  \Phi_{(n-3,2)}
-
{1 \over 2} \,\Phi_{(n-2,0)} \right\} \\
F_{(n-2,0)} &=& {n(n-1) \over 12(n+2)!} \, \Phi_{(n-2,0)}
\end{eqnarray*}

\section{Coefficients $\lambda_{ij}(k)$}

Coefficients $\lambda_{ij}(k)$ (see Eq.(\ref{degen-integrals}))

\begin{eqnarray*}
\lambda_{11}(k)=1-{4k^3+8k^2+19k+16\over2(2k^2+k+1)}\left({4\over9}\right)^k
+{48k^4-32k^3-126k^2+47k+119\over
2k^2+k+1}\left({1\over9}\right)^k
\end{eqnarray*}
\begin{eqnarray*}
\lambda_{22}(k)=1-{4k^4+77k^2+9k+12\over4(2k^2+k+1)}\left({4\over9}\right)^k
+{88k^4-200k^3+288k^2-75k+3\over 2k^2+k+1}\left({1\over9}\right)^k
\end{eqnarray*}
\begin{eqnarray*}
\lambda_{12}(k)={\sqrt{2k(2k-1)(k+1)(2k+3)}\over4(2k^2+k+1)}\left\{(7-6k)\left({4\over9}\right)^k
-8(6k^2-25k+20)\left({1\over9}\right)^k\right\}
\end{eqnarray*}

\end{document}